\documentclass[journal]{IEEEtran}

\usepackage{xcolor,soul,framed} 

\colorlet{shadecolor}{yellow}
\usepackage[pdftex]{graphicx}
\graphicspath{{../pdf/}{../jpeg/}}
\DeclareGraphicsExtensions{.pdf,.jpeg,.png}

\usepackage[cmex10]{amsmath}
\usepackage{array}
\usepackage{graphicx}
\usepackage{mdwmath}
\usepackage{mdwtab}
\usepackage{eqparbox}
\usepackage{subfigure}
\usepackage{multirow}
\usepackage{cite}
\usepackage{url}
\usepackage{hyperref}
\usepackage{amssymb}
\hypersetup{
	colorlinks=true,
	linkcolor=blue,
	filecolor=blue,      
	urlcolor=red,
	citecolor=blue,
}

\newcommand{\etal}{\textit{et al}.}
\newcommand{\ie}{\textit{i}.\textit{e}.}
\newcommand{\eg}{\textit{e}.\textit{g}.}
\newcommand{\etc}{\textit{etc}}

\begin{document}
\bstctlcite{IEEEexample:BSTcontrol}
    \title{PUGAN: Physical Model-Guided Underwater Image Enhancement Using GAN with Dual-Discriminators}
  \author{Runmin Cong,~\IEEEmembership{Senior Member,~IEEE,}
      Wenyu Yang, Wei Zhang,~\IEEEmembership{Senior Member,~IEEE,} Chongyi Li,~\IEEEmembership{Senior Member,~IEEE,} Chun-Le Guo, Qingming Huang,~\IEEEmembership{Fellow,~IEEE,}
      and~Sam Kwong,~\IEEEmembership{Fellow,~IEEE}


\thanks{Runmin Cong is with the Institute of Information Science, Beijing Jiaotong University, Beijing 100044, China, also with the School of Control Science and Engineering, Shandong University, Jinan 250061, China, and also with the Key Laboratory of Machine Intelligence and System Control, Ministry of Education, Jinan 250061, China (e-mail: rmcong@sdu.edu.cn).}
\thanks{Wenyu Yang is with the Institute of Information Science, Beijing Jiaotong University, Beijing 100044, China, also with the Beijing Key Laboratory of Advanced Information Science and Network Technology, Beijing 100044, China (e-mail: wyuyang@bjtu.edu.cn).}
\thanks{Wei Zhang is with the School of Control Science and Engineering, Shandong University, Jinan 250061, China, and also with the Key Laboratory of Machine Intelligence and System Control, Ministry of Education, Jinan 250061, China (e-mail: davidzhang@sdu.edu.cn).}
\thanks{Chongyi Li and Chun-Le Guo are with the College of Computer Science, Nankai University, Tianjin 300350, China  (e-mail: lichongyi25@gmail.com; guochunle@nankai.edu.cn).}
\thanks{Qingming Huang is with the School of Computer Science and Technology, University of Chinese Academy of Sciences, Beijing 101408, China, also with the Key Laboratory of Intelligent Information Processing, Institute of Computing Technology, Chinese Academy of Sciences, Beijing 100190, China, and also with Peng Cheng Laboratory, Shenzhen 518055, China (email: qmhuang@ucas.ac.cn).}
\thanks{Sam Kwong is with the Department of Computer Science, City University of Hong Kong, Hong Kong SAR, China, and also with the City University of Hong Kong Shenzhen Research Institute, Shenzhen 51800, China (e-mail: cssamk@cityu.edu.hk).}

}


\maketitle

\begin{abstract}
Due to the light absorption and scattering induced by the water medium, underwater images usually suffer from some degradation problems, such as low contrast, color distortion, and blurring details, which aggravate the difficulty of downstream underwater understanding tasks. Therefore, how to obtain clear and visually pleasant images has become a common concern of people, and the task of underwater image enhancement (UIE) has also emerged as the times require. Among existing UIE methods, Generative Adversarial Networks (GANs) based methods perform well in visual aesthetics, while the physical model-based methods have better scene adaptability. Inheriting the advantages of the above two types of models, we propose a physical model-guided GAN model for UIE in this paper, referred to as PUGAN. The entire network is under the GAN architecture. On the one hand, we design a Parameters Estimation subnetwork (Par-subnet) to learn the parameters for physical model inversion, and use the generated color enhancement image as auxiliary information for the Two-Stream Interaction Enhancement sub-network (TSIE-subnet). Meanwhile, we design a Degradation Quantization (DQ) module in TSIE-subnet to quantize scene degradation, thereby achieving reinforcing enhancement of key regions. On the other hand, we design the Dual-Discriminators for the style-content adversarial constraint, promoting the authenticity and visual aesthetics of the results. Extensive experiments on three benchmark datasets demonstrate that our PUGAN outperforms state-of-the-art methods in both qualitative and quantitative metrics. The code and results can be found from the link of \url{https://rmcong.github.io/proj\_PUGAN.html}.
\end{abstract}


%
\IEEEpeerreviewmaketitle


\section{Introduction}

\IEEEPARstart{O}{cean} contains extremely rich resources, and researchers can use remotely operated underwater vehicles (ROVs) \cite{tao2017image} to collect images and videos from the underwater environment for further perception and exploitation.
Due to the complex underwater environments and lighting conditions, underwater images are degraded by wavelength-dependent absorption and scattering (\ie, forward scattering and back-scattering) \cite{li2019underwater}. 
As thus, it becomes very challenging to directly obtain valuable information from degraded underwater images, which hinders the further development of other ocean-related tasks \cite{hambarde2021uw}. Therefore, the research on underwater image enhancement (UIE) technology emerges as the times require, which has important practical application value.

\begin{figure}[!]
\centering
\subfigure{
\begin{minipage}[t]{1\linewidth}
\centering
\includegraphics[width=9cm,height=3cm]{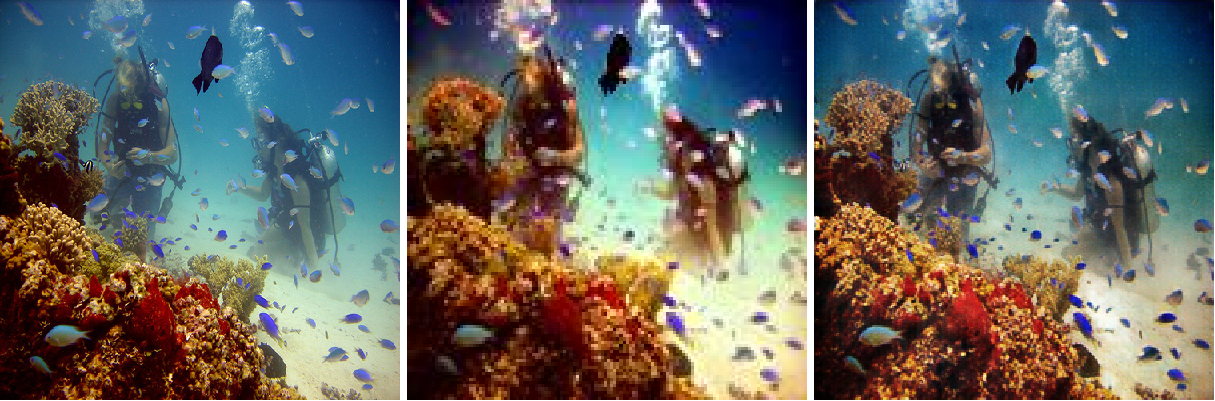}
{\qquad\qquad\qquad\qquad\qquad\qquad\qquad (a) \qquad\qquad\qquad\quad(b) \qquad\qquad\qquad\quad(c)}
\end{minipage}
}%

\subfigure{
\begin{minipage}[t]{1\linewidth}
\centering
\includegraphics[width=9cm,height=3cm]{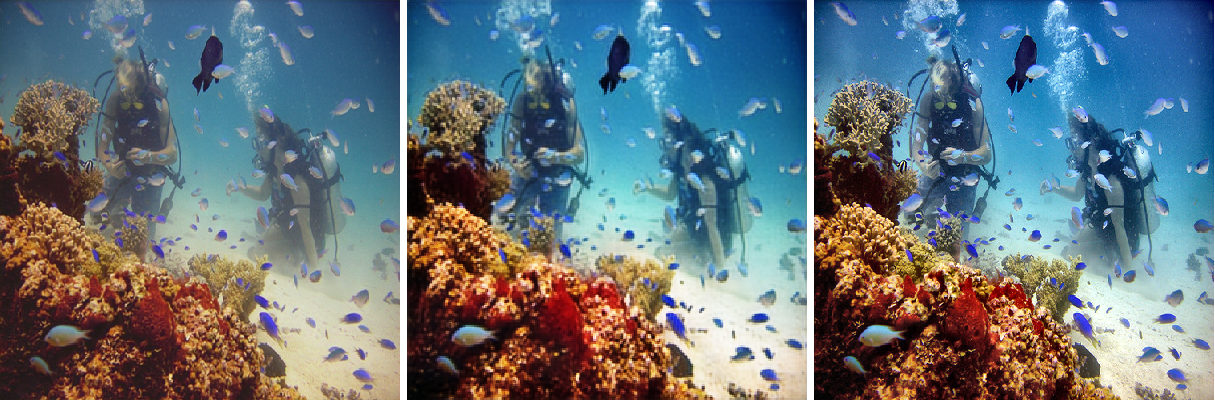}
{(d) \qquad\qquad\qquad\quad(e) \qquad\qquad\qquad\quad(f)}
\end{minipage}
}%
\centering
\caption{Samples of different UIE methods. (a) Original underwater image. (b)-(e) The enhancement results generated by  GDCP \cite{peng2018generalization}, FUnIE-GAN \cite{islam2020fast}, Ucolor \cite{li2021underwater}, PUGAN (ours). (f) Ground truth. }
\label{1}
\label{fig1}
\end{figure}

The degradation of underwater images is mainly manifested in the color distortion caused by the absorption effect of water, and blurring caused by the scattering effect of suspended particles in water (\ie, organic particles, planktonic microorganisms, \etc.) \cite{9825662}.
Due to the uniqueness and complexity of underwater imaging process, the enhancement methods designed for other degeneration scenes (\eg, low-light scene \cite{guo2020zero,li2021learning,crm/tits22/low-light} or foggy scene \cite{9447190, 9252912,crm/mtap22/dehazing,crm/tmm20/dehazing}) do not generalize well to the UIE task.
Moreover, even some methods specially designed for UIE task are not satisfactory in enhancement quality.
In the early days of this task, the traditional methods prevailed, which can be roughly divided into the non-physical model based methods and physical model based methods. 
The non-physical model methods mainly focus on the adjustment of image pixels, such as dynamic pixel range stretching \cite{iqbal2010enhancing}, pixel distribution adjustment \cite{ghani2015underwater}, and image fusion \cite{ancuti2017color,gao2019underwater,ancuti2012enhancing}. 
These methods heavily rely on hand-crafted feature designs, which makes them prone to over- or under-enhancement, affecting the overall visual effect. 
By contrast, the physical model-based methods utilize some priors and assumptions to model the process of underwater optical imaging \cite{he2010single,galdran2015automatic, drews2016underwater, crm/JEI16/underwater, li2016underwater,zhang2022underwater}, thereby generating the clean image through model inversion.
Although the physical model cannot fully and realistically simulate the underwater imaging process, it is undeniable that modeling the underwater imaging process is conducive to solving the unique visual problems of underwater images, such as color distortion.
In recent years, deep learning technology has also greatly promoted the development of the computer vision task, including object detection \cite{crm/tcyb22/glnet,crm/acmmm21/CDINet,crm/nips20/CoADNet,crm/tetci22/PSNet,crm/tmm22/TNet,crm/tip22/CIRNet,crm/tnnls22/360SOD,crm/tcsvt22/weaklySOD}, medical analysis \cite{crm/jbhi22/polyp,crm/tce22/covid,crm/tip20/MCMT-GAN,crm/tim22/covid}, remote sensing interpretation\cite{crm/tcyb22/rsi,crm/tip21/DAFNet,crm/tgrs22/RRNet,crm/tgrs19/rsi}, content enhancement \cite{crm/ijcai20/SR,crm/acmmm21/bridgenet,crm/mtap22/dehazing,crm/tmm20/dehazing}, video generation \cite{zhang2023controlvideo} and underwater image enhancement \cite{li2018emerging,li2017watergan,li2020underwater}. 
Designing deep UIE networks in an end-to-end manner is the most common practice, with the Convolutional Neural Networks (CNN)-based and Generative Adversarial Networks (GAN)-based framework as the main genre.
Among them, the GANs-based methods \cite{islam2020fast,chen2018deep,ignatov2017dslr,zhu2017unpaired} have achieved surprising performance in improving perceptual image quality from an extensive collection of paired or unpaired data. 
The learning-based method takes advantage of the powerful learning ability of the network and can achieve good results in some scenarios. However, underwater environments are often complex and diverse, and purely relying on network learning may distort the enhancement results. As shwon in Fig. \ref{fig1}, we provide some visual examples of different enhancement methods, including the traditional GDCP  method\cite{peng2018generalization}, the GAN-based FUnIE-GAN  method\cite{islam2020fast}, the CNN-based Ucolor method \cite{li2021underwater}, and our method. It can be seen that the given comparison algorithms either have noticeable color deviations or the results are blurry. 
In fact, the learning model can be guided and supplemented by the domain knowledge of underwater imaging to achieve better enhancement effects. 
Here, the domain knowledge of underwater imaging refers to the modeling of the imaging principle of the underwater image, that is, the reason of defects such as color distortion, blurring, and detail distortion. 
For example, the transmission map in the underwater physical model \cite{galdran2015automatic} contains factors describing the quality of underwater imaging in different regions, which can guide the model inversion and degradation modeling. 



Through the above analysis, GANs have strong learning ability but lack generalization ability \cite{yang2022generalization}, and modeling the underwater imaging process is beneficial to understand the characteristics of underwater images.
Therefore, we hope to design a network architecture that can effectively combine them to give play to complementary advantages and collaborative promotion.
To this end, we perform network design under a GANs architecture, aiming to learn the mapping between the original and cleaned underwater images by training on the real underwater datasets.
For the generator, we fully combine the physical model and the CNN-based model, thereby forming the Physical Model-Guided Generator (Phy-G).
On the one hand, we adopt the commonly used image degradation model to simulate the underwater physical distortion, and design a learning based network, named Parameters Estimation subnetwork (Par-subnet), to estimate the parameters for physical model inversion, mainly including the transmission map and attenuation coefficient. With these parameters, a color enhanced underwater image can be generated by inversion.
On the other hand, considering that the color enhanced image has good generalization properties and can be further used as auxiliary information to guide the CNN-based enhancement network, we design a UNet-like structure with two input branches, named Two-Stream Interaction Enhancement sub-network (TSIE-subnet).
Specifically, with the help of color enhanced image and estimated transmission map, we design a Degradation Quantization (DQ) module to locate and quantify the distortion degree of the scene, thereby enabling targeted encoder feature filtering and reinforcing.
Furthermore, training with synthetic datasets is difficult to simulate real-world situations due to the lack of absolute supervision with real underwater images as references, which may severely degrade the performance of deep learning algorithms. To better constrain the underwater images generated by the generator to be close to real and clear underwater images, in addition to the pixel-level loss and perceptual loss, we design a novel Dual-Discriminators (Dual-D) structure to judge the reconstruction results of the generator, following a style-content synergy mechanism. Among them, one discriminator is used to judge the overall style of the result, and the other is to concatenate the result with the estimated depth map to judge the authenticity of its image content. 
The Par-subnet and TSIE-subnet in the Phy-G cooperate with each other and are unified in the network framework with GAN as the main body.


The main contributions are highlighted as follows:
\begin{itemize}
\item[(1)] Considering the respective advantages of the physical model and the GAN model for the UIE task, we propose a Physical Model-Guided framework using GAN with Dual-Discriminators (PUGAN), consisting of a Phy-G and a Dual-D. Extensive experiments on three benchmark datasets demonstrate that our PUGAN outperforms state-of-the-art methods in both qualitative and quantitative metrics.

\item[(2)] We design two subnetworks in the Phy-G, including the Par-subnet and the TSIE-subnet, for the parameter estimation of physical model and the physical model-guided CNN-based enhancement, respectively. On the one hand, we introduce an intermediate variable in the Par-subnet, \ie, depth, to enable effective estimation of the transmission map. On the other hand, we propose a DQ module in TSIE-subnet to quantify the distortion degrees and achieve targeted encoder feature reinforcing.

\item[(3)] In addition to the pixel-level global similarly loss and perceptual loss, we design the style-content adversarial loss in the Dual-D to constrain the style and content of the enhanced underwater image to be realistic.
\end{itemize}

The rest of the paper is organized as follows. Section II reviews the related work. Section III presents the details of the proposed PUGAN framework. The experimental comparisons and discussions are shown in Section IV. Finally, the conclusion is drawn in Section V.


\section{Related Work}


\subsection{Traditional UIE methods} 
In the early days, underwater image enhancement methods mainly adjusted the pixel values of images according to the characteristics of underwater images such as color distortion and low contrast.
This adjustment is generally performed in the spatial domain, such as dynamic pixel range stretching \cite{iqbal2010enhancing}, pixel distribution adjustment \cite{ghani2015underwater}, and image fusion \cite{ancuti2017color,gao2019underwater,ancuti2012enhancing}. Song \etal \cite{song2021enhancement} presented a comprehensive underwater visual reconstruction paradigm that comprises three procedures, \ie, the E-procedure, the R-procedure, and the H-procedure. Ancuti \etal \cite{ancuti2019color} proposed a color channel compensation (3C) pre-processing method. Moreover, as a pre-processing step, the 3C operator can improve traditional restoration methods.
In addition, many traditional UIE methods are based on physical model, which focus on estimating all parameters in the physical model and inverting the clear underwater images \cite{crm/JEI16/underwater}. These methods usually rely on some priors, such as red channel prior \cite{zhang2022underwater}, underwater dark channel prior \cite{galdran2015automatic,crm/JEI16/underwater}, minimum information prior \cite{li2016underwater}, general dark channel prior \cite{he2010single,peng2018generalization}, and blurriness prior \cite{drews2016underwater}. 

Due to the limited expression ability of hand-crafted features or priors used by traditional UIE methods, their enhancement effect and adaptability to different water environments are unsatisfactory. For example, we noticed that the color correction algorithms in \cite{lin2020attenuation,zhang2022underwater} do not always work well with some challenging underwater images. 
For this reason, learning-based methods have gradually become the mainstream of research hotspots in recent years, especially after entering the era of deep learning.


\subsection{Learning-based UIE method} 

In recent years, deep learning has been widely used in underwater image enhancement, which has dramatically improved the effect of enhancement \cite{li2020underwater,li2019underwater,crm/spl21/underwater,9774330,li2017watergan}. Li \etal \cite{li2020underwater} proposed to simulate the realistic underwater image according to different water types. With ten types of synthesized underwater images, ten underwater image enhancement (UWCNN) models were trained and each UWCNN model was used to enhance the corresponding type. Li \etal \cite{li2019underwater} proposed the WaterNet for real UIE task by fusing the inputs with the predicted confidence maps to achieve the image enhancement. Deep learning networks often need to learn distribution from a large number of paired data. 
But for underwater image enhancement task, high-quality paired training data is often difficult to obtain. 
To this end, some researchers introduced the Generative Adversarial Networks (GANs) in the UIE task by using a two-player min-max game between the generator (usually based on CNNs) and the discriminator. 
In the final equilibrium state, the generator can effectively learn to model the underlying distribution. Unlike clear supervision, the discriminator loss tends to determine the distribution and style of the input, and is more suitable for image enhancement tasks that map from low-quality domains to high-quality domains.
Li \etal \cite{li2017watergan} proposed a GAN-based UIE method, which is trained by synthetic underwater image and outputs the depth map at the same time. 
Jiang \etal \cite{9774330} designed a global-local discriminative structure for the UIE task. These methods hardly take into account the physical model, but due to the unique characteristics of the underwater image, which may lead to ineffectiveness in the face of a harsh underwater environment.

Recently, some methods have embedded physical models into deep learning networks. 
Li \etal \cite{li2021underwater} proposed a medium transmission-guided encoder network to force the network to emphasize the quality-degraded regions.
Wang \etal \cite{wang2021joint} proposed a joint iterative network for UIE, which divides the enhancement into the color correction sub-task and dehazing sub-task, corresponding to light absorption and scattering effects, respectively.
Hambarde \etal \cite{hambarde2021uw} proposed the UW-GAN to estimate the depth, and then use it to restore a high-quality underwater image. Due to the introduction of the physical model, the performance of the enhancement model and the generalization were significantly improved. 

Our method fully incorporates physical model in a learned manner under a GAN architecture. 
Different from previous methods, we physically invert the color enhanced image in a learned manner, and use it as guidance information to participate in the secondary enhancement under the CNN architecture, in which a DQ module is also designed to quantify the degradation degree and achieve differentiated region enhancement.
Besides, the style-content adversarial loss in the Dual-D is provided by two independent discriminators, hoping to promote visual aesthetics and content authenticity, respectively.


	

\begin{figure*}[!t]
\centering
\centerline{\includegraphics[width=1\textwidth]{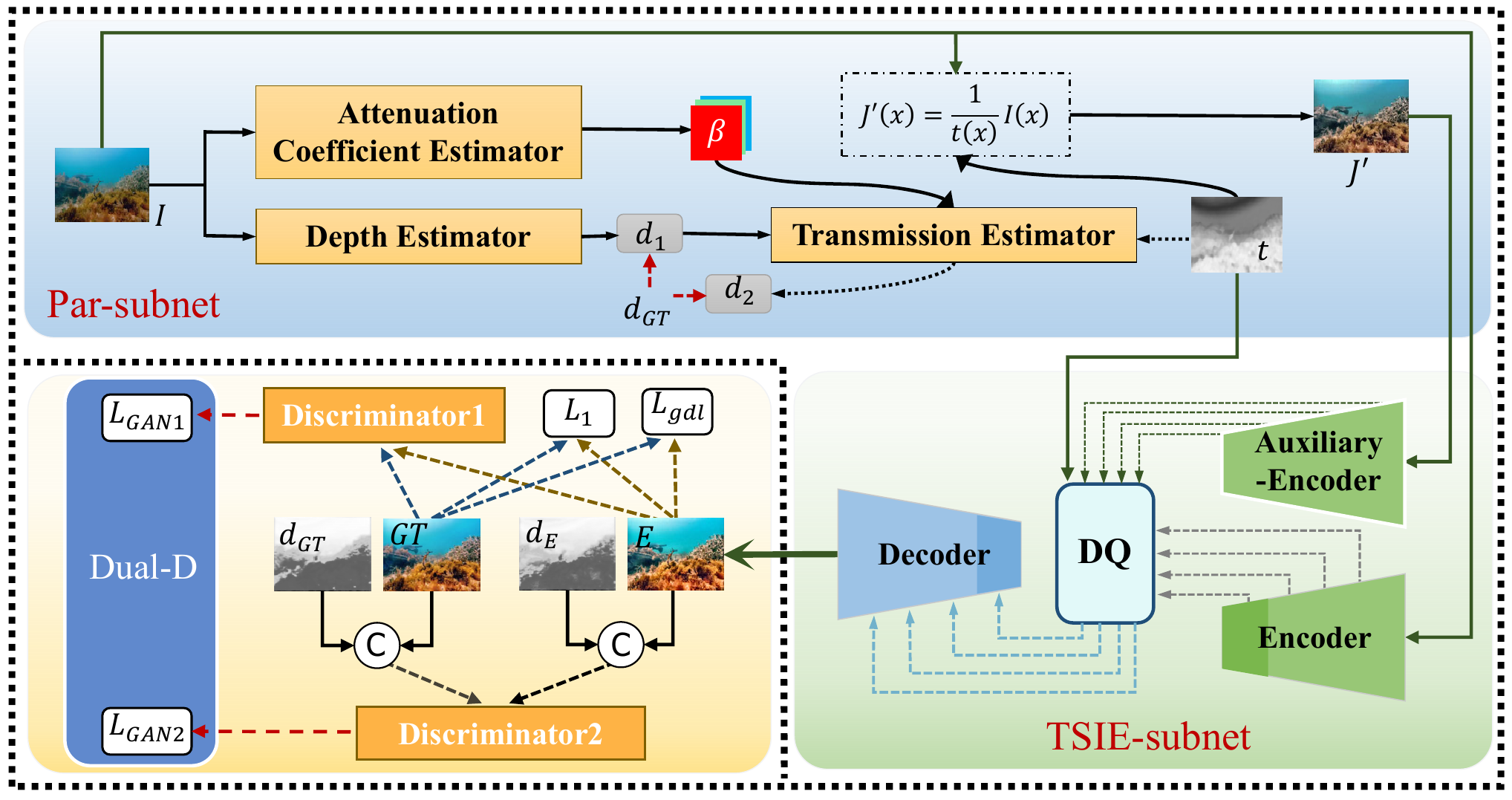}}
\caption{Overview of the proposed PUGAN for UIE task, including a Phy-G and a Dual-D under the GAN architecture. In the Phy-G, the Par-subnet is used to estimate the physical parameters (\eg, transmission map $t$ and attenuation coefficient $\beta$) required for restoring a color-enhanced image $J^{'}$. The TSIE-subnet aims to achieve the CNN-based end-to-end enhancement, where a degradation quantization (DQ) module is used to quantify the distortion degree of the scene, thereby guiding and generating the final enhanced underwater image $E$. The objective function consists of four parts, including global similarity loss ${L_1}$, perceptual loss ${L_{gdl}}$, style adversarial loss ${L_{GAN_1}}$, and content adversarial loss ${L_{GAN_2}}$.
		}
	\label{fig2}
\end{figure*}

\section{Proposed Method}

For the underwater image enhancement task, we design a PUGAN to learn the mapping from the degraded underwater image $I$ to the enhanced underwater image $E$, as shown in Fig. \ref{fig2}. The pipeline of the entire network is under the GAN architecture, including a Physical Model-Guided Generator (Phy-G) and a Dual-Discriminators (Dual-D). 
In the Phy-G part, we first propose a parameters estimation subnetwork (Par-subnet) to model the physical imaging process of underwater images, which estimates the physical parameters required for restoring a color-enhanced image $J^{'}$, including the transmission map and attenuation coefficient.
Then, with the help of the color-enhanced image $J^{'}$ and transmission map, we design a two-stream interaction enhancement subnetwork (TSIE-subnet) to achieve the CNN-based end-to-end enhancement, where a degradation quantization (DQ) module located between the encoder and the decoder is used to quantify the distortion degree of the scene, thereby achieving the regional and differential encoder feature reinforcement, and generating the final enhanced underwater image $E$. 
We set up Dual-D to guarantee the visual aesthetics and authenticity of the enhanced results. The enhanced image $E$ and the corresponding ground truth are fed into the Discriminator 1 to constrain the overall style of the results, while incorporating them with the depth map into Discriminator2 to constrain the authenticity of the structural content.
In the following subsections, we will detail the critical part of our model, including the Phy-G, Dual-D, and loss functions. 


\subsection{Physical Model-Guided Generator}
In our PUGAN, the Phy-G is a two-stage UIE network for generating enhanced underwater images, consisting of a Par-subnet and a TSIE-subnet. In the following subsubsections, we will introduce them one by one.

\subsubsection{Parameters Estimation Subnetwork (Par-subnet)}

For the image enhancement tasks, physical modeling is a more common practice, such as the well-known atmospheric scattering models in the dehazing task \cite{9018379, 9088248}. In fact, in order to simulate the degradation process of underwater images, people also try to physically model the underwater scenes.
Let's first review the physical model of the underwater imaging process, which can be expressed as \cite{galdran2015automatic}:
\begin{equation}
I(x) = J(x)t(x) + A(1 - t(x)),
\label{eq1}
\end{equation}
where $I$ is the observed underwater image, $J$ denotes the restored image, $A$ represents the background light, and $t$ is the transmission map, describing the portion of the light that is not scattered and reaches the camera, which can be further expressed as:
\begin{equation}
t(x) = e^{ - \beta d(x)},
\label{eq2}
\end{equation}
where $\beta$ is the attenuation coefficient of the water, and $d$ is the depth of scene. This equation indicates that the scene radiance is attenuated exponentially as the depth. Therefore, the depth can also reflect the attenuation of the scene to a certain extent. 

From Eq. (\ref{eq1}), we can inversely derive the calculation formula of the enhanced image $J$ as:
\begin{equation}
J(x) = \frac{1}{{t(x)}}I(x) - A(\frac{1}{{t(x)}} - 1).
\label{eq3}
\end{equation}
The first term of this formula is mainly used to correct the color of $I$, and the second term is to remove the influence of background light. In fact, in underwater images, color distortion is a key factor in our subjective visual quality. Therefore, we focus on obtaining color-corrected underwater images through physical model inversion during the first-stage enhancement process of the network. In this way, we can invert the color-enhanced image $J^{'}$ according to Eq. (\ref{eq3}) from the original image $I$:
\begin{equation}
J^{'}(x) = \frac{1}{{t(x)}}I(x).
\label{eq4}
\end{equation}


Obtaining underwater images through model inversion has good interpretability and scene adaptability, and is also the key to traditional underwater image enhancement methods. But how to estimate the parameters of the physical model is a formidable challenge. Some previous works \cite{peng2018generalization,li2021underwater} used prior knowledge to estimate parameters involved in physical models. However, the prior knowledge is usually less robust and may lead to severe estimation bias in challenging underwater scenarios. 
To sum up, considering the advantages of physical models and the challenges of parameter estimation, on the one hand, we introduce a physical enhancement mode to improve the interpretability of the overall framework, and on the other hand, we estimate the parameters required by the model in a learning manner to improve its accuracy.

\begin{figure}[t]
	 \centering	
	\centerline{\includegraphics[width=0.5\textwidth]{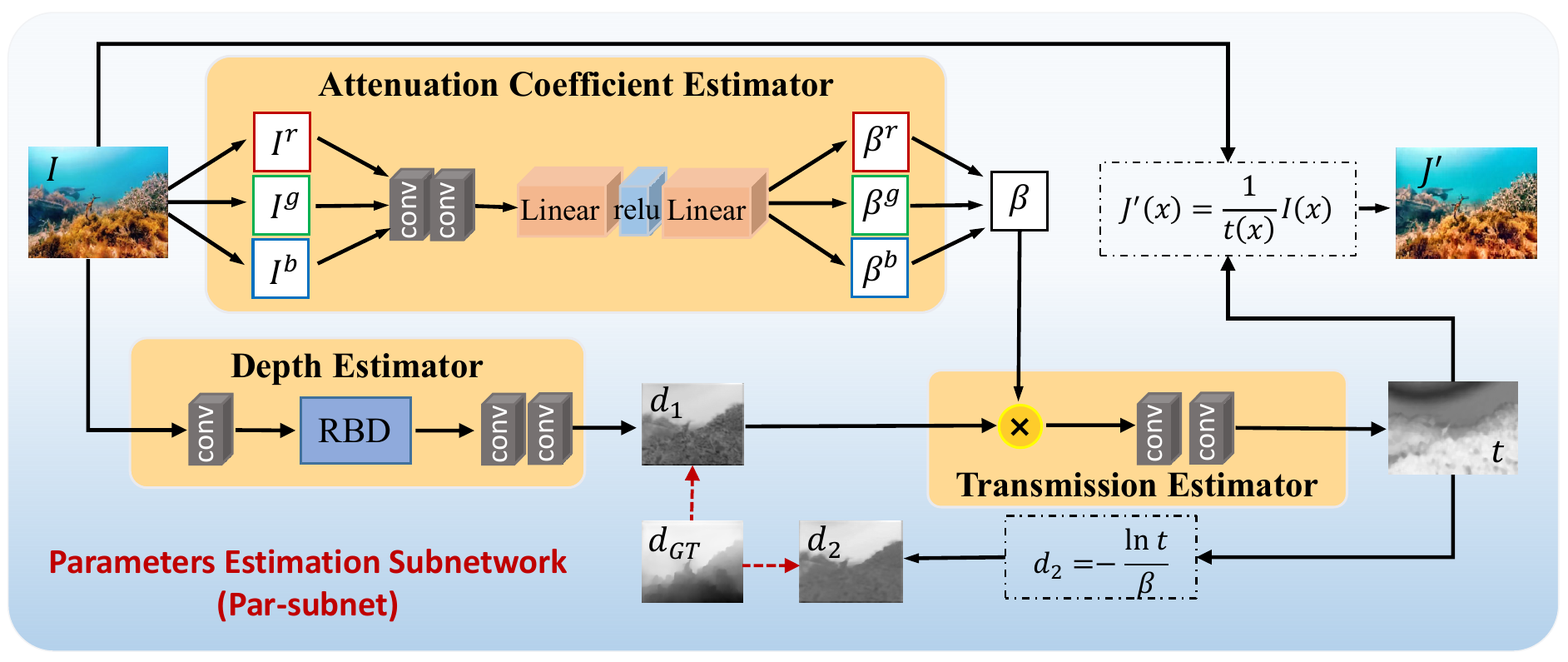}}
	\caption{The schematic illustration of Par-subnet. It mainly includes three modules, namely Depth Estimator, Attenuation Coefficient Estimator, and Transmission Estimator. The results of the Depth Estimator and Attention Coefficient Estimator are used to estimate the transmission map. 
	Finally, the estimated transmission map and the original image are used to restore the color-enhanced underwater image through the model inversion.
		}
	\label{fig3}
\end{figure}

To this end, we design a Par-subnet to estimate the parameters in the physical model, which is shown in Fig. \ref{fig3}. 
According to Eq. (\ref{eq2}) and Eq. (\ref{eq4}), the parameters we need to estimate mainly include the attenuation coefficient $\beta$ and the depth map $d$, which are used to compute the transmission map $t$.
First, we can use the attenuation coefficient estimator and the depth estimator to calculate the attenuation coefficient and depth map. Specifically, we feed each color channel of the original underwater image into a convolutional block, generating three attenuation coefficients, which can be formulated as:
\begin{equation}
{\beta ^c} = {linear} (relu({linear} ({conv.p.r}(I^c)))),
\label{eq6}
\end{equation}
where $I^c$ is the $c$ channel of the original underwater image, $c=\{r,g,b\}$ indexes the color channel,  $conv.p.r$ includes two convolutional layers with the kernel size of $3 \times 3$ followed by a pooling layer and a ReLU activation $relu$, and ${linear}$ denotes the linear layer. Then, these three channels are concatenated as the final attenuation coefficient:
\begin{equation}
\beta  = cat({\beta ^r},{\beta ^g},{\beta ^b}),
\label{eq7}
\end{equation}
where $cat$ is a channel-wise concatenation.

In the lower left branch, we use the RBD block \cite{islam2020simultaneous} to directly estimate the depth map $d_1$, which can be expressed as:
\begin{equation}
d_1 = \sigma(conv(conv.b.r(\texttt{RBD}(conv.b.r(I))))),
\label{d1}
\end{equation}
where $conv.b.r$ denotes the convolutional layer with the kernel size of $3 \times 3$ followed by a normalization layer and a ReLU activation, $\sigma$ denotes the Sigmoid function, $conv$ is the convolutional layer, and $\texttt{RBD}$ represents the RBD block \cite{islam2020simultaneous}. 

Following Eq. (\ref{eq2}), the transmission map has an exponential relationship with the product of depth and attenuation coefficient. So, we design a network to estimate the transmission map $t$ with the $({d_1}\cdot{\beta)}$ as input:
\begin{equation}
t = \sigma(conv(conv.b.r({d_1}\cdot{\beta}))).
\label{eq9}
\end{equation}

So far, with estimated transmission map $t$ and attenuation coefficient $\beta$, we can obtain a color-enhanced underwater image $J^{'}$ according to the physical model formulated in Eq. (\ref{eq4}), which will play an important role in the TSIE-subnet. 
It should be noted that, in order to ensure the quality of the transmission map, we use the attenuation coefficient again to calculate the depth map as follows:
\begin{equation}
d_2 =  - \frac{{\ln t}}{\beta }.
\label{eq10}
\end{equation}
In this way, we can further strengthen the constraints on the estimated transmission map from the perspective of physical model inversion, thereby improving the accuracy of the entire parameter estimation network. The whole structure of the Par-subnet is relatively simple, mainly as an auxiliary structure before starting to enhance the underwater image.

\subsubsection {Two-Stream Interaction Enhancement Subnetwork (TSIE-subnet)}

In the first stage, we invert color-enhanced underwater images with better interpretability using the learned physical model parameters. But as mentioned before, the enhancement effect is not perfect due to the exclusion of background light.
Therefore, we re-enhance the underwater images under the CNN network architecture in the second stage guided by the color-enhanced images, thereby forming a two-stream architecture to realize the interaction of multi-source information.
Under this two-stream structure, how to achieve sufficient information interaction is the key problem to be solved.
It is well known that underwater image enhancement needs to correct serious color distortion and eliminate blurring and duskiness.
In other words, we should pay more attention to degradation recovery while maintaining the integrity of image information.

\begin{figure}[t]
	 \centering
	\includegraphics[width=0.48\textwidth]{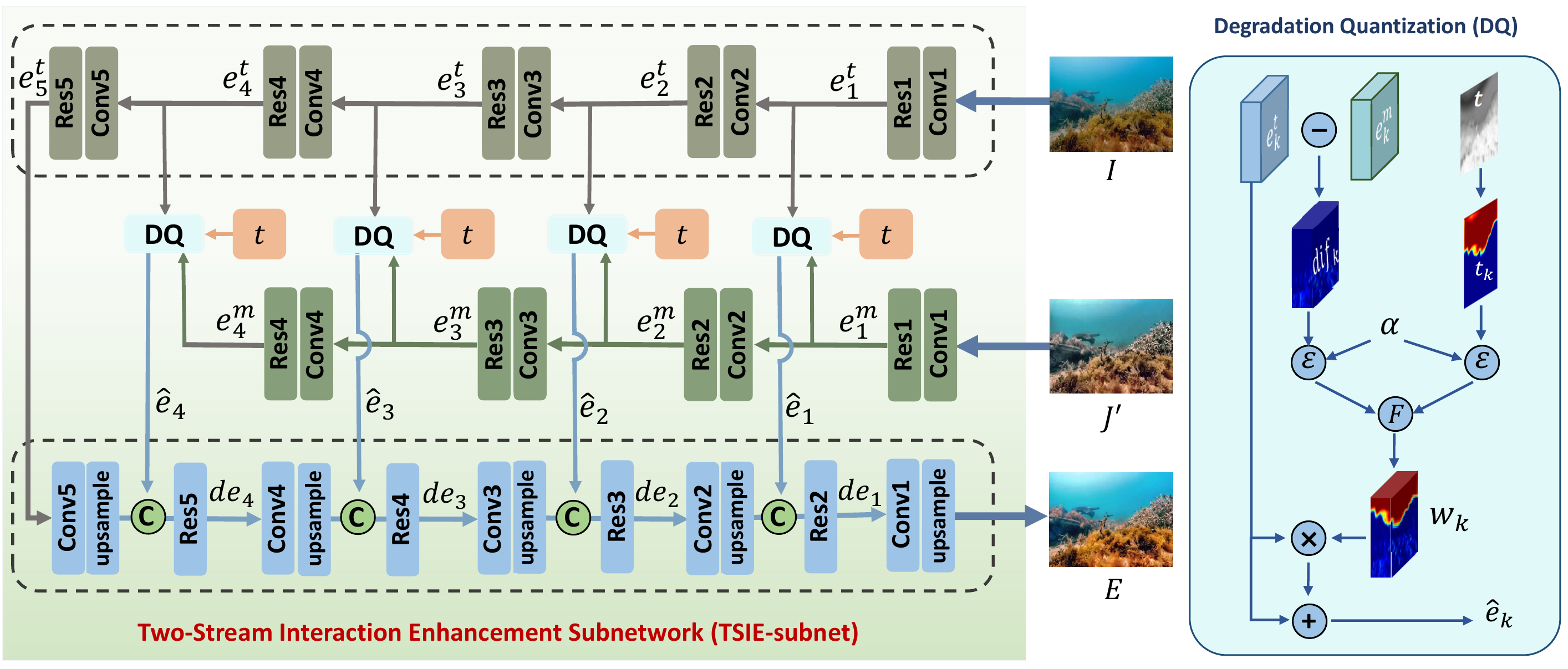}		
	\caption{The schematic illustration of TSIE-subnet, following an encoder-decoder structure, in which the two-stream encoder features are transferred to the corresponding decoder layer after passing through the DQ module. The right side of this figure provides the detailed structure of the DQ module.
		}
	\label{fig4}
\end{figure}

Different regions in the underwater scene have different degrees of degradation. If the same level of enhancement is performed indiscriminately, the optimal enhancement effect will often not be achieved.
Let's look back at the underwater physical imaging model, the degree of image degradation is negatively related to the transmission rate (the value of each pixel in the transmission map represents the transmission rate at that position, between 0 and 1). Thus, based on this prior, we identify some regions that are prone to degradation from the transmission map.
Moreover, the difference between the color-enhanced image $J^{'}$ and the original degraded underwater image $I$ can directly reflect the degradation of underwater image quality to a certain extent, and then can give the TSIE-subnet a clear optimization direction.
Therefore, we adaptively localize the degradation in the scene with the help of the transmission map $t$ and the color-enhanced underwater image $J^{'}$ to enable the targeted encoder feature reinforcement. 
To this end, combining the above two cases, we design a DQ module to adaptively locate and quantify these seriously degraded regions in the scene, thereby guiding the encoder features fusion and reinforcement. The schematic illustration of the TSIE-subnet is shown in Fig. \ref{fig4}, which is a two-stream encoder and one-stream decoder structure.

First, we feed the original underwater image $I  \in  {\mathbb{R}^{3\times256\times256}}$ and the color enhanced underwater image $J^{'}  \in  {\mathbb{R}^{3\times256\times256}}$ into the top and middle encoders to extract multi-level features, denoted as $e_k^t$ and $e_k^m$, respectively. The encoder contains five convolution-residual blocks, where each block consists of a convolutional layer with the kernel size of $3\times3$ followed by a pooling layer and ReLu activation, and a residual layer. 
Then, we utilize the DQ module to locate the seriously degraded regions and fuse the encoder features.
As analyzed earlier, we can locate the degraded regions in the scene with the help of the color-enhanced map and the transmission map, respectively.

On the one hand, we can locate severely degraded regions by directly comparing the difference between the color-enhanced image features and the original image features, which can be described as:
\begin{equation}
dif_k = conv.b.r(|e_k^t - e_k^m|)\cdot \varepsilon(conv.b.r(|e_k^t - e_k^m|)-\alpha),
\label{eq11}
\end{equation}
where $dif_k$ is the feature difference with reference to the color-enhanced features, indicating how much information the image needs to be supplemented, $\varepsilon(\cdot)$ denotes the step function, and $\alpha$ is a threshold, which is set to 0.7 in experiments. 



On the other hand, the degree of degradation of underwater images is negatively correlated with the transmission characteristics.
Therefore, we can also identify some regions that are prone to degradation from the transmission map:
\begin{equation}
t_k = (1-maxpool(t))\cdot \varepsilon(1-maxpool(t)-\alpha),
\label{eq13}
\end{equation}
where $maxpool$ is the maxpooling operation. In fact, the $t_k$ has the same size as $e_k^t$. Moreover, a larger value of $t_k$ means more severe degeneration, and then the enhancement of these regions in the decoding stage needs to be strengthened.

Combining these two aspects, the final weights can be defined as follows:
\begin{equation}
{w_k} = \sigma(conv((conv.b.r({t_k} + di{f_k})))).
\label{eq15}
\end{equation}

Subsequently, these weights are applied to the input features $e_k$ to generate the updated features ${\hat e}_k$ through the residual connection:
\begin{equation}
{\hat e_k} = {e_k^t} + {e_k^t} \otimes {w_k},
\label{eq16}
\end{equation}
where $\otimes$ is Hadamard product. The severely degraded areas in ${\hat e}_k$ are given larger weights, so that the subsequent networks can focus on strengthening these regions.

Finally, we implement feature decoding using convolution-residual blocks symmetric to the encoder, including $3 \times 3$ convolution layer, upsampling layer, and ReLU activation. The decoder features can be formulated as:
\begin{equation}
{de_k} = res(cat[conv.b.r(up({de_{k+1}})),{\hat e_k}]),
\label{eq17}
\end{equation}
where $up$ is the upsampling operation, $res$ denotes the residual operation, and $de_{k+1}$ represents the decoder features of the $k+1$ layer. We operate on $de_1$ using a convolutional layer to obtain the final result $E$.

\subsection{Dual-Discriminators (Dual-D)}

For the GAN-based UIE methods, the generator aims to obtain the final enhanced underwater image, and the discriminator focuses on judging whether the generated image is real or fake. If the generated image fails to fool the discriminator, the generator needs further training and optimization. The whole process can be constrained by an adversarial loss  \cite{li2020deep}, which can be described as:
\begin{equation}
\begin{array}{l}
{\arg \mathop {\min }\limits_G \mathop {\max }\limits_D L_{GAN}}(G,D) \\
 = {{\mathbb{E}}_{\{X,Y\}}}[\log D(Y)] + {{\mathbb{E}}_{\{X,Y\}}}[\log (1 - D(G(X|Z)))]
\end{array}
\label{eq18}
\end{equation}
where $X$ and $Y$ denote the source domain (low-quality image) and desired domain (enhanced image), $G$ represents the generator aiming to learn a mapping of $X \rightarrow Y$, and $D$ denotes the discriminator.


	

To better constrain the generated underwater images to be close to real and clear underwater images, we design a novel Dual-Discriminators (Dual-D) structure to judge the reconstruction results of the generator, following a style-content synergy mechanism. Unlike the existing GAN-based UIE methods\cite{chen2018deep,ignatov2017dslr,zhu2017unpaired}, our Dual-D includes two discriminators to jointly constrain the style and content of the generated images. First, Discriminator1 is the standard usage of the discriminator to judge whether the overall style is authentic, and it does not pay attention to how much a certain area in the image should be enhanced, as long as it looks like a high-quality underwater image. This may cause some areas to be over-enhanced, or some important areas are not sufficiently enhanced. In fact, from the perspective of the human visual system, people’s perception of content in different regions is different. For example, people will pay more attention to the foreground area in underwater images, \ie, the regions with small depth values. If different regions are treated equally, over-enhancement may occur in distant regions, which is inconsistent with human aesthetics. Therefore, the Discriminator2 considers the depth structure of the scene and constraints TSIE-subnet to be able to perform a discriminative enhancement. The two discriminators complement each other to implement constraints on the style and content of the enhanced results.

For the structure selection of discriminators, we prefer to divide the image into several patches instead of focusing on the whole image. This is mainly because, on the global image, the discriminator may only score from the overall style such as blur, color balance, contrast, \etc, which is not what we expect. We hope that the network can improve the image quality to different degrees according to different image contents. Thus, we adopt the Markovian PatchGAN as discriminators, which only penalizes the patch-scale structure. The discriminator attempts to classify whether each $N\times N$ patch in the image is real or fake, which helps capture high-frequency features such as local texture and style. Moreover, the PatchGAN has fewer parameters and runs faster\cite{li2020deep}. Specifically, we feed the image with the size of $256\times 256\times c$ into four convolution layers, get the output with the size of $16\times 16\times 1$, and calculate the average value as the final response of the discriminator. In each convolution layer, the kernel size is set to $3\times 3$ with the stride of 2, and the use of the nonlinear layer and normalization layer is the same as in TSIE-subnet.


Based on the above, the style-content adversarial loss in our Dual-D structure can be represented as:
\begin{small}
\begin{equation}
\begin{array}{l}
{\arg \mathop {\min }\limits_G \mathop {\max }\limits_{D_1} L_{GAN_1}}(G,D_1) \\
 = {{\mathbb{E}}_{\{I,Y\}}}[\log D_{1}(Y)] + {{\rm E}_{\{I,Y\}}}[\log (1 - D_{1}(E))],
 \end{array}
 \label{eq19}
\end{equation}
\end{small}
\begin{small}
\begin{equation}
\begin{array}{l}
{\arg \mathop {\min }\limits_G \mathop {\max }\limits_{D_2} L_{GAN_2}}(G,D_2) \\
 = {{\mathbb{E}}_{\{I,Y,d\}}}[\log D_{2}(Y,d_Y)]
 + {{\mathbb{E}}_{\{I,Y,d\}}}[\log (1 - D_{2}(E,d_E))],
 \end{array}
 \label{eq20}
\end{equation}
\end{small}
where $D_1$ and $D_2$ denote the Discriminator1 and Discriminator2, $E$ is the final generated enhancement image, $Y$ denotes the ground truth of the enhancement image, $d_E$ and $d_Y$ are the estimated depth map from the generated enhancement image $E$ and real enhancement image $Y$, respectively.





\subsection{Training Strategy and Loss Function}\label{sec-train}

The generator of our proposed PUGAN includes Par-subnet and TSIE-subnet, and Dual-D as the discriminators.
For network training, we employ a two-stage training strategy. 
In the first stage, the Par-subnet is pre-trained offline.
Then, in the second stage, we train the whole PUGAN, where the parameters of the Par-subnet are fixed, and the TSIE-subnet and Dual-D are trained alternately.


\subsubsection{\textbf{Par-subnet training}}
We use the synthetic dataset \cite{li2020underwater} to train the Par-subnet, including five types of open ocean water and five types of coastal water. There are corresponding labels for the depth map and attenuation coefficient for each water type. 
We randomly select $200$ images from the synthetic dataset for training. In the training phase, each stage of our model is trained separately for 60 epochs with a batch size of $4$. The learning rate is fixed to $1{e^{ - 4}}$. We first train the attenuation coefficient estimator and then freeze their parameters to train the depth estimator and transmission estimator. 
To control the accuracy of the transmission map, we use the transmission map and attenuation coefficient to compute the depth map again. Therefore, the loss of Par-subnet is defined as follows: 
\begin{equation}
\begin{array}{l}
{L_p} = \frac{1}{{H\times W}}[\sum\limits_{m = 1}^H {\sum\limits_{n = 1}^W {(|d(m,n) - {d_1}(m,n)|) + } }\\
\sum\limits_{m = 1}^H {\sum\limits_{n = 1}^W {(|d(m,n) - {d_2}(m,n)|)] + } } \frac{1}{3}\sum\limits_{c = 1}^3 {(|{{\hat \beta } ^c} - {\beta ^c}|)} 
\end{array}
\label{eq21}
\end{equation}
where $d_1$ and $d_2$ are the estimated depth map, respectively, $d$ is the ground truth of the depth map, ${\hat \beta }$ and ${\beta }$ denote the ground truth and estimated value of the attenuation coefficient, respectively.



\subsubsection{\textbf{PUGAN training}}

After the Par-subnet training is completed, we perform the second-stage training on the entire PUGAN structure.
Following the settings in \cite{li2021underwater}, the training data include $800$ pairs of underwater images selected from the UIEB dataset \cite{li2019underwater} and $1250$ synthetic underwater images selected from the synthesized UIE dataset \cite{isola2017image}. 

In order to make the generated image as visually pleasing as possible while maintaining its authenticity of the image, we use global similar loss, perceptual loss, and adversarial loss to compose the final loss function:
\begin{equation}
\begin{array}{l}
L = {\lambda _1}\cdot\mathop {\arg }\limits_{} \mathop {\min }\limits_G \mathop {\max }\limits_{D_1}  {L_{GAN1}}(G,{D_1})\\
 + {\lambda _2}\cdot\mathop {\arg }\limits_{} \mathop {\min }\limits_G \mathop {\max }\limits_{D_2}  {L_{GAN2}}(G,{D_2})\\
 + {\lambda _3}\cdot{L_1}(E,Y) + {\lambda _4}\cdot{L_{con}}(E,Y)
\end{array}
\label{eq22}
\end{equation}
where ${\lambda _1}$, ${\lambda _2}$, ${\lambda _3}$ and ${\lambda _4}$ are scaling factors that represent the contributions of respective loss components, ${L_1}$ is the global similarly loss \cite{9646904}, ${L_{gdl}}$ is the perceptual loss \cite{islam2020fast}, and ${L_{GAN_1}}$ and ${L_{GAN_2}}$ are style and content adversarial losses, respectively.

\section{Experiment}

\subsection{Implementation Details}


We independently evaluate the performance on widely used real UIE benchmark datasets with corresponding high-quality ground truth, including UIEB dataset \cite{li2019underwater}, UFO-120 dataset \cite{islam2020fast}, and EUVP dataset \cite{islam2020fast}. Each dataset is divided into a training dataset and a testing dataset. 
The UIEB dataset consists of 950 underwater images (890 images with ground truth and 60 challenging images without ground truth), we take 800 pairs of underwater images to train our PUGAN, and the remaining 90 pairs are used for testing. Following the settings in \cite{li2021underwater}, we incorporate 1250 synthetic underwater images selected from the synthesized underwater image dataset \cite{isola2017image} to expand the training set. 
The EUVP dataset is a large-scale underwater taken under various visibility, containing 11435 pairs of images for training and 1030 pairs of images for testing. 
The UFO-120 dataset \cite{islam2020fast} contains  1500 annotated samples for large-scale training, and an additional 120 testing samples.
We name their testing sets Test-UIEB, Test-UFO, and Test-EUVP, respectively.

We implement our model using the PyTorch toolbox with an NVIDIA GeForce RTX 3090 GPU. We also implement our network by using the MindSpore Lite tool\footnote{\url{https://www.mindspore.cn/}}. The model parameters are initialized with a normal distribution.  The first-stage training of the Par-subnet is introduced in Section \ref{sec-train}. In the second-stage training of the PUGAN, the initial learning rate is set to $1{e^{ - 3}}$ and divided by 10 every 50 epochs. We train 200 epochs with a batch size of 16. All input images are resized to $256\times 256$. 
For quantitative evaluation, following \cite{li2019underwater}, the Peak Signal-to-Noise Ratio (PSNR) and Mean-Square Error (MSE) are introduced. A higher PSNR or a lower MSE score denotes that the result is closer to the reference image in terms of image content or structure.
In addition, we also use some non-reference metrics for evaluation, including UIQM \cite{panetta2015human}, FDUM \cite{yang2021reference}, UCIQE \cite{7300447}, and CCF \cite{wang2018imaging}.
They quantify the quality of underwater images in terms of colorfulness, sharpness, contrast, fog density \etc. In an ideal state, the higher the scores on these non-reference metrics, the better the image quality.

\subsection{Comparison with State-of-the-Art Methods}
To demonstrate the effectiveness of our proposed method, we conduct qualitative and quantitative comparisons with some state-of-the-art UIE methods on the above three datasets. The comparison methods includes 6 non-learning-based methods (\ie, GDCP \cite{peng2018generalization}, ACDE \cite{zhang2022underwater}, HLRP \cite{9854113}, MLLE \cite{zhang2022underwater11}, UNTV \cite{9548907}, and SPDF \cite{9895452}) and 8 deep learning-based methods (\ie, deep-sesr \cite{islam2020simultaneous}, FUnIE-GAN \cite{islam2020fast}, WaterNet \cite{li2019underwater}, UWCNN \cite{li2020underwater}, JI-Net \cite{wang2021joint}, ACPAB \cite{lin2020attenuation}, TOPAL 
\cite{9774330}, and Ucolor 
\cite{li2021underwater}). It should be noted, WaterNet \cite{li2019underwater}, FUnIE-GAN \cite{islam2020fast}, and TOPAL 
\cite{9774330} are also GAN-based methods.


\subsubsection{Quantitative Evaluation}


We provide the average PSNR and MSE values on the Test-UIEB, Test-UFO, and Test-EUVP datasets in Table \ref{tab2}.
It is observed that the performance of traditional methods is often not comparable to deep learning-based methods due to the limited representational power of hand-crafted features. 
For example, even the latest papers published on IEEE TIP 2022 (\ie, HLRP \cite{9854113} and MLLE \cite{zhang2022underwater11}) still fail to make the top three.
In contrast, the deep learning-based methods have clear performance advantages, among which our method comes out on top in terms of PSNR and MSE on three testing datasets. 
Specifically, compared with the \textbf{second-best competitor} for PSNR metric, our method achieves a percentage gain of 5.1\% on the Test-UIEB dataset, 2.1\% on the Test-UFO dataset, and 2.2\% on the Test-EUVP dataset, respectively.
For the MSE metric, the \textbf{minimum percentage gain} of our method reaches 30.8\% on the Test-UIEB dataset, 15.8\% on the Test-UFO dataset, and 2.9\% on the Test-EUVP dataset, respectively.
These all demonstrate the strength and effectiveness of our model.

\begin{table}[t]
\renewcommand\arraystretch{1.3}
			\caption{The evaluations of different methods on three datasets in terms of average PSNR (DB) and MSE ($\times 10^{3}$) values. The top three results are marked in red, blue, and green, respectively.}
\scalebox{0.89}{
\begin{tabular}{c|cc|cc|cc}\hline
Datasets             & \multicolumn{2}{c|}{Test-UIEB}                                                  & \multicolumn{2}{c|}{Test-UFO}                                                   & \multicolumn{2}{c}{Test-EUVP}                                                    \\\hline
Methods              & PSNR$\uparrow$                                  & MSE$\downarrow$                                 & PSNR$\uparrow$                                  & MSE$\downarrow$                                  & PSNR$\uparrow$                                 & MSE$\downarrow$                                  \\\hline
GDCP \cite{peng2018generalization}        & 13.72                                 & 3.37                                 & 14.33                                 & 2.87                                 & 13.35                                  & 3.58                                  \\
ACDE \cite{zhang2022underwater}        & 16.85                                 & 1.67                                 & 14.31                                 & 2.83                                 & 15.03                                  & 2.35                                  \\
HLRP \cite{9854113}        & 12.17                                 & 4.24                                 & 11.69                                 & 4.66                                 & 11.32                                  & 5.08                                  \\
MLLE \cite{zhang2022underwater11}        & 18.82                                 & 1.12                                 & 15.05                                 & 2.45                                 & 15.06                                  & 2.32                                  \\
UNTV \cite{9548907}        & 16.57                                 & 1.88                                 & 17.12                                 & 1.42                                 & 17.50                                  & 1.39                                  \\
SPDF \cite{9895452}        & {\color[HTML]{00B050} \textbf{19.85}}                                 & {\color[HTML]{00B050} \textbf{0.92}}                                 & 17.57                                 & 1.37                                 & 18.82                                  & 1.09                                  \\
\hline
deep-sesr \cite{islam2020simultaneous}   & 15.77                                 & 2.08                                 & {\color[HTML]{0070C0} \textbf{23.22}} & {\color[HTML]{0070C0} \textbf{0.38}} & {\color[HTML]{00B050} \textbf{23.22}}                                 & {\color[HTML]{0070C0} \textbf{0.35}}                                  \\
FUnIE-GAN \cite{islam2020fast}   & 18.07                                 & 1.78                                 & {\color[HTML]{00B050} \textbf{22.97}} & {\color[HTML]{00B050} \textbf{0.41}} & {\color[HTML]{0070C0} \textbf{23.53}}  & {\color[HTML]{00B050} \textbf{0.41}}  \\
WaterNet \cite{li2019underwater}    & 19.81     & 1.02     & 19.63                                 & 0.83                                 & 20.58                                  & 0.71                                  \\
UWCNN \cite{li2020underwater}       & 13.26                                 & 4.00                                 & 16.41                                 & 1.98                                 & 17.72                                  & 1.40                                  \\
JI-Net \cite{wang2021joint} & 18.21                                 & 2.46                                 & 16.54                                 & 1.78                                 & --                                      & --                                     \\
ACPAB \cite{lin2020attenuation}       & 15.20                                 & 2.52                                 & 17.04                                 & 1.73                                 & 18.06                                  & 1.40                                  \\
TOPAL \cite{9774330}        & {\color[HTML]{00B050} \textbf{19.85}}                                 & 0.93                                 & 19.31                                 & 0.83                                 & 19.98                                  & 0.75                                  \\
Ucolor \cite{li2021underwater}      & {\color[HTML]{0070C0} \textbf{20.61}} & {\color[HTML]{0070C0} \textbf{0.78}} & 19.45                                 & 0.85                                 & 20.08                                  &  0.76           \\
\hline
PUGAN                & {\color[HTML]{FF0000} \textbf{21.67}} & {\color[HTML]{FF0000} \textbf{0.54}} & {\color[HTML]{FF0000} \textbf{23.70}} & {\color[HTML]{FF0000} \textbf{0.32}} & {\color[HTML]{FF0000} \textbf{24.05}} & {\color[HTML]{FF0000} \textbf{0.34}}\\
\hline
\end{tabular}
}
\label{tab2}
\end{table}

\begin{table*}[t]
\renewcommand\arraystretch{1.3}
			\caption{The evaluations of different methods on three datasets in terms of UIQM, FDUM, UICQE, and CCF metrics. The top three results are marked in red, blue, and green, respectively.}
\scalebox{1}{
\begin{tabular}{c|cccc|cccc|cccc}\hline
Datasets             & \multicolumn{4}{c|}{Test-UIEB}                                                  & \multicolumn{4}{c|}{Test-UFO}                                                   & \multicolumn{4}{c}{Test-EUVP}                                                    \\\hline
Methods              & UIQM$\uparrow$                                   & FDUM$\uparrow$            & UICQE$\uparrow$                                 & CCF$\uparrow$
& UIQM$\uparrow$                                  & FDUM$\uparrow$ 
& UICQE$\uparrow$                                 & CCF$\uparrow$
& UIQM$\uparrow$                                  & FDUM$\uparrow$ 
& UICQE$\uparrow$                                 & CCF$\uparrow$
\\\hline
input                & 2.69                                 & 0.36                   &0.52
&19.59        & 2.48                                 & 0.48            &0.56           &30.03                     & 2.49                                 & 0.45                              &0.55        &30.27   \\
Ground truth         & 3.01                                 & 0.55                  &{\color[HTML]{0070C0} \textbf{0.62}}        &27.34           & 2.88                                 & 0.67                     &0.60       &28.53            & 2.88                                 & 0.62                     &0.58       
  &31.11            \\\hline
GDCP \cite{peng2018generalization}        & 2.67  & {\color[HTML]{FF0000} \textbf{0.84}}       
 &{\color[HTML]{00B050} \textbf{0.61}}        &{\color[HTML]{0070C0} \textbf{47.28}}         & 2.10                             & {\color[HTML]{FF0000} \textbf{0.81}}     &{\color[HTML]{0070C0} \textbf{0.66}}          &{\color[HTML]{0070C0} \textbf{62.83}}           & 2.43  & {\color[HTML]{FF0000} \textbf{0.87}} &{\color[HTML]{0070C0} \textbf{0.63}}             &{\color[HTML]{00B050}\textbf{57.92}}  
 \\
ACDE \cite{zhang2022underwater}        & {\color[HTML]{FF0000} \textbf{3.41}}                                 & 0.49               &0.56                    &29.05                              & {\color[HTML]{FF0000} \textbf{3.35}}                                 & 0.51            &0.57  
            &33.44                                 & {\color[HTML]{FF0000} \textbf{3.30}}                                & 0.43          &0.56             &33.38                       \\
HLRP \cite{9854113}   & 1.99                                 & {\color[HTML]{0070C0} \textbf{0.81}}                &{\color[HTML]{FF0000} \textbf{0.66}}                   &{\color[HTML]{FF0000} \textbf{55.25}}                 & 2.47                                 & {\color[HTML]{FF0000} \textbf{0.81}}                &{\color[HTML]{FF0000} \textbf{0.67}}                &{\color[HTML]{FF0000} \textbf{63.23}}                   & 2.41                                 & {\color[HTML]{00B050} \textbf{0.75}}         &{\color[HTML]{FF0000} \textbf{0.65}}                   &{\color[HTML]{FF0000} \textbf{64.56}}                        \\
MLLE \cite{zhang2022underwater11}   & 2.65                                 & 0.66               &{\color[HTML]{00B050} \textbf{0.61}}               &{\color[HTML]{00B050} \textbf{40.12}}                       & 2.39                                 & {\color[HTML]{00B050} \textbf{0.76}}   &{\color[HTML]{00B050} \textbf{0.62}}                   &{\color[HTML]{00B050} \textbf{56.43}}                             & 2.28                                & 0.69              &0.61               &{\color[HTML]{0070C0} \textbf{60.31}}                   \\
UNTV \cite{9548907}        & 2.94                                 & {\color[HTML]{00B050}\textbf{0.72}}          &0.59                &26.37                       & 2.60                                 & {\color[HTML]{0070C0} \textbf{0.80}}   &{\color[HTML]{00B050} \textbf{0.62}}              &38.81                              & 2.47                                  & {\color[HTML]{0070C0} \textbf{0.77}}               &{\color[HTML]{00B050} \textbf{0.62}}	       &40.78                   \\
SPDF \cite{9895452}        & 3.08                                 & 0.44                        &0.56	                &17.46         & {\color[HTML]{0070C0} \textbf{3.18}}                                 & 0.50            &0.56	                  &22.96                      & {\color[HTML]{00B050} \textbf{3.19}}                                  & 0.27            &0.55	      &24.54                      \\
\hline
deep-sesr \cite{islam2020simultaneous}   & 2.97                                 & 0.41          &0.53	                   &15.97                       & 3.07                                 & 0.61                     &0.59	                 &23.90                & 3.10                                & 0.54                   &0.57	               &24.34              \\
FUnIE-GAN \cite{islam2020fast}   & {\color[HTML]{0070C0} \textbf{3.34}}                                 &0.68                 &0.56	                &21.38               & 2.97                                & 0.58       &0.60	           &27.85              & 2.99                                 & 0.56 
&0.59               	&30.10 \\
WaterNet \cite{li2019underwater}    & 3.04                                 & 0.44               &0.58	                  &16.68                  & 3.08                                 & 0.53               &0.59             	&25.60                  & 3.06                                 & 0.50               &0.58	            &27.17                  \\
UWCNN \cite{li2020underwater}       & 2.21                                 & 0.28               
  &0.48	          &10.65                  & 2.93                                 & 0.28               &0.52	        &15.91                  & 2.96                                 & 0.39 &0.52	          &19.02                                \\
JI-Net \cite{wang2021joint} & 2.67                                  & 0.57                     &0.59	     &25.98            & {\color[HTML]{00B050} \textbf{3.17}}  & 0.54                    &0.59	        &28.70             & {\color[HTML]{0070C0} \textbf{3.24}}  & 0.67 
&0.58	&27.38    \\
ACPAB \cite{lin2020attenuation}       & 2.92                                & 0.56              &0.58	             &33.66                   & 3.06                                 & 0.51             &0.58	        &33.78                    & 2.98                             & 0.45                &0.58	           &35.90                 \\
TOPAL \cite{9774330}        & 3.08                                 & 0.48                       &0.57	               &22.82          & 3.02                                 & 0.36                         &0.61	              &28.85        & 3.01                                  & 0.32               &0.43	                &28.50                   \\
Ucolor \cite{li2021underwater}      & {\color[HTML]{00B050} \textbf{3.30}}                                 & 0.43            &0.57	        &17.65                     & 3.14                    & 0.52                    &0.59	              &24.53              & 3.12                                  & 0.49         &0.58	          &26.51                        \\\hline
PUGAN                & 3.28 & 0.68 
 &{\color[HTML]{0070C0} \textbf{0.62}}	&27.94          & 2.85                                 & 0.64     &0.60	&33.49         & 2.94                            & 0.53   
      &0.60	              &30.34         \\
\hline
\end{tabular}
}
\label{tab1}
\end{table*}

\begin{figure*}[!t]
	 \centering
	\includegraphics[width=0.98\textwidth]{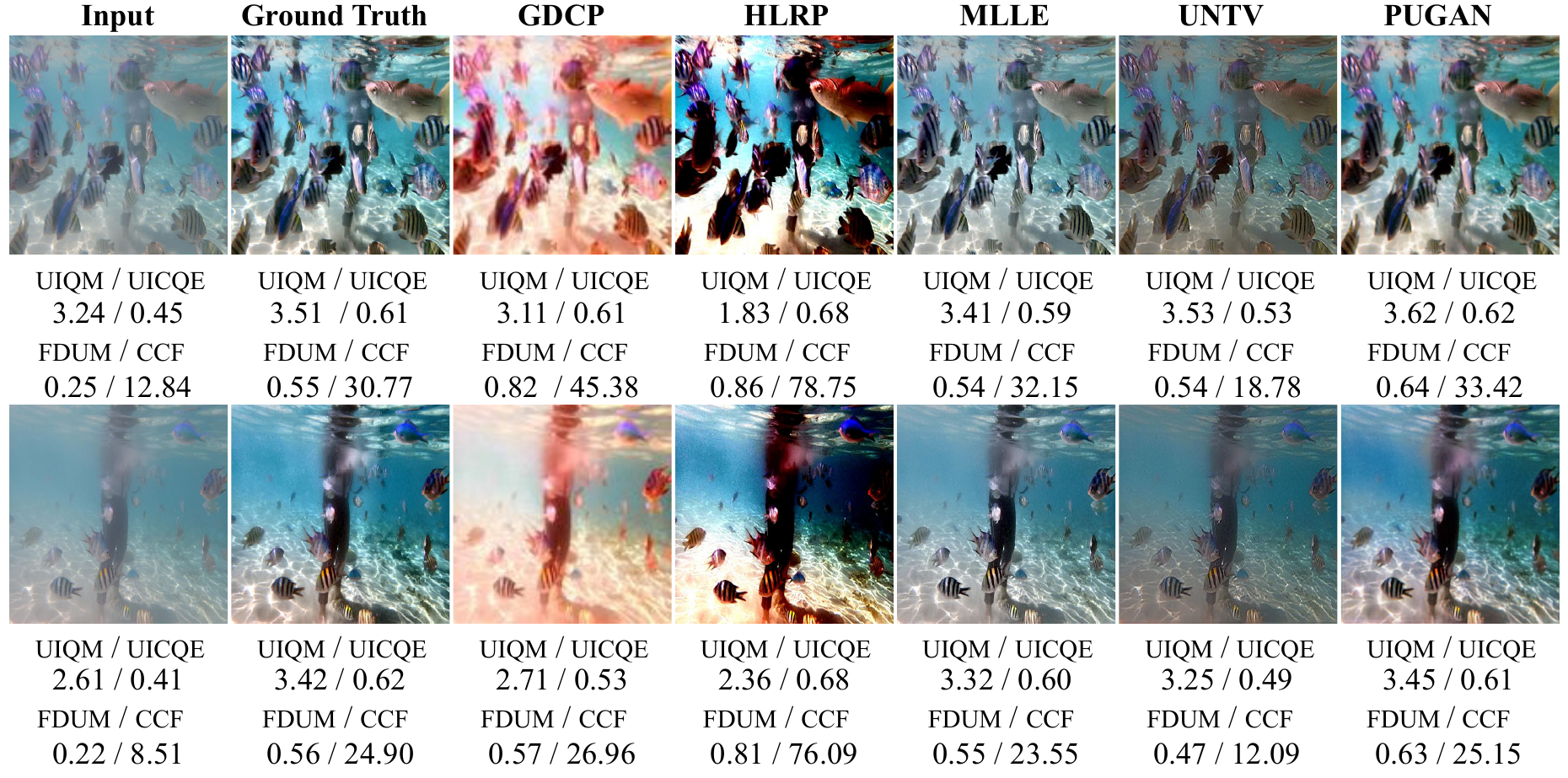}	
	
	\caption{The scores of non-reference metrics are displayed under each visualization result.
		}
	\label{fig4.1}
\end{figure*}


\begin{figure*}[!h]
\centering
\centerline{\includegraphics[width=0.86\textwidth]{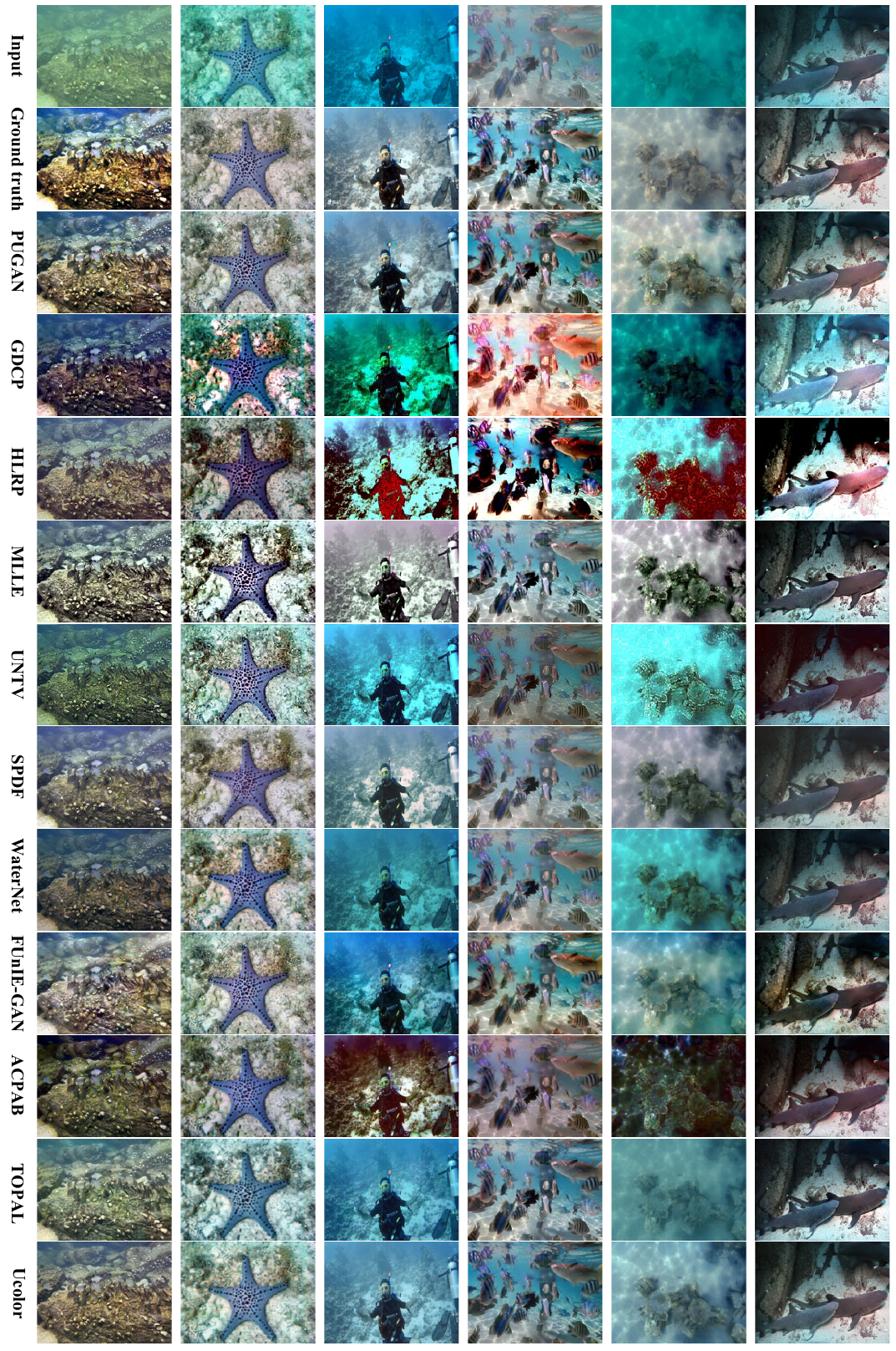}}
\caption{Visualizations of different comparison algorithms.}
	\label{fig5}
\end{figure*}

In addition, we report the average UIQM, FDUM, UICQE, and CCF scores of different methods on three datasets in Table \ref{tab1}. For UIQM metric, ACDE \cite{zhang2022underwater} achieves the best performance on all three datasets. Similarly, GDCP \cite{peng2018generalization} wins the victory in terms of the FDUM metric on these three datasets. HLRP \cite{9854113} performs best in UICQE and CCF metrics on all datasets, and obtains the same FDUM score as GDCP \cite{peng2018generalization} on the Test-UFO dataset.
Overall, the deep learning methods have no advantage for these no-reference evaluation metrics, and our PUGAN only ranks second in terms of the UICQE metric on the Test-UIEB dataset.
We further select some visualization results of three traditional algorithms that scored high on the non-reference metrics and compare them with the original underwater image, ground truth, and our method in Fig. \ref{fig4.1}. We can see that the scores of HLRP \cite{9854113} are much higher than other algorithms in terms of UICQE, FDUM, and CCF metrics, but the visual effect is far inferior to some other algorithms, \ie, the color of the object appears serious deviation. Likewise, the enhancement results of GDCP \cite{peng2018generalization} also exhibit obvious color distortion, but also achieve high scores on the non-reference metrics. 
Moreover, it can be seen from Table \ref{tab1} that GT does not score as well as some algorithms on non-reference metrics. Specifically, GT tends to score lower than traditional algorithms on the non-reference indicators, which is determined by the calculation methods of these indicators and the properties of traditional enhancement algorithms. Judging from the design of non-reference indicator calculation, although the measurement standards and combination methods of the four indicators are different, almost all of them focus on color and contrast. As such, these metrics tend to give higher scores to images with high color saturation, brightness, and contrast. But sometimes this is not consistent with people's subjective visual experience.
For example, as shown in the first image of Fig. \ref{fig4.1}, the UICQE score is lower than that of HLRP, FUDM and CCF scores are both lower than that of GDCP and HLRP. But the fact is that the visual quality of GT is significantly better than that of GDCP and HLRP methods. It also shows to some extent that non-reference evaluation metrics are not always reliable.
The above problems have also been pointed out in some existing works. Jiang \etal \cite{9774330} pointed out that although some methods have defects such as color distortion and overbrightness, they can still achieve high scores on the UICQE metric. Chen \etal \cite{9646904} also argued that UIQM and UICQE metrics highlight the bright and high-contrast features without considering the color shifts and artifacts, resulting in some over-enhanced underwater images also achieving high scores on these no-reference metrics. From another point of view, there is still a lot of research space for the no-reference underwater image quality assessment. In addition, the guidance of no-reference metrics can also be considered when constructing a new UIE dataset in the future, thereby jointly promoting the common development of the field of underwater content understanding.
Therefore, we can pay more attention to the metrics of PSNR and MSE to judge the advantages of different methods.




\subsubsection{Qualitative Evaluation}
Some visual experimental results of different methods are shown in Fig. \ref{fig5}. As can be seen, 
our method is closer to the ground truth in terms of color, brightness, definition, \etc, and has advantages in the following aspects:

\textbf{Color distortion correction.} Color distortion is arguably one of the most common phenomena of underwater image degradation, which is caused by the different light absorption capacities of water at different wavelengths. Since color distortion can greatly affect visual aesthetics, one of the goals of enhancement algorithms is color correction.
For example, the second and fifth images in Fig. \ref{fig5} are cases with greenish distortion. 
It can be seen that, except for our method, MLLE \cite{zhang2022underwater11}, and SPDF \cite{9895452}, the enhancement effect of other comparison methods still suffers from greenish color distortion. In particular, GDCP \cite{peng2018generalization} method further exacerbates the color distortion around the image, and the deep learning-based methods (\ie, FUnIE-GAN \cite{islam2020fast} and Ucolor \cite{li2021underwater}) also has a very obvious green effect. In contrast, our PUGAN produces more natural, realistic colors with relatively high definition. The advantage of our method in handling similar scenarios can also be reconfirmed from the first and fifth images. 
In addition to greenish distortion, bluish distortion is also a common case, as shown in the third image. We can see that only Ucolor \cite{li2021underwater} method and our PUGAN can effectively correct the bluish distortion. While for other methods, the results are less satisfactory. For example, GDCP \cite{peng2018generalization} method introduces a severe green cast, HLRP \cite{9854113} method makes the image reddish, and MLLE \cite{zhang2022underwater11} method almost removes the color of the image.

\textbf{Complex scenes.} 
For the complex and cluttered scene in the fourth image, our method still achieves better performance. The fish in this image are characterized by a large number, rich colors, and varying depths, which undoubtedly increases the difficulty of restoration. GDCP \cite{peng2018generalization}, HLRP \cite{9854113}, and ACPAB \cite{lin2020attenuation} methods make the enhanced color of fish wrong. And the enhanced results of SPDF \cite{9895452}, WaterNet \cite{li2019underwater}, and Ucolor \cite{li2021underwater} methods are blurry and hazy.
By contrast, the result of our PUGAN is closer to ground truth in terms of definition and color, where each fish and background are properly adjusted.


\textbf{Low contrast and Low light.} Forward scattering tends to cause low contrast of underwater images, as shown in the second and fourth images, and it is often necessary to enlarge color contrast while removing blur. From the results, ACPAB \cite{lin2020attenuation} and GDCP \cite{peng2018generalization} methods have insufficient ability to process low contrast images, and some methods even bring additional distortion (\eg, HLRP \cite{9854113}, UNTV \cite{9548907}, and WaterNet \cite{li2019underwater}).
Furthermore, underwater images are sometimes affected by insufficient lighting, which requires UIE algorithms to increase brightness. As shown in the last image, some traditional methods (\eg, GDCP \cite{peng2018generalization} and HLRP \cite{9854113}) over-enhance and make the image too bright. And some deep learning methods either fail to improve brightness sufficiently (\eg, ACPAB \cite{lin2020attenuation} and WaterNet \cite{li2019underwater}) or introduce color distortion (\eg, FUnIE-GAN \cite{islam2020fast}). 
In contrast, our method achieves good results on both low-contrast enhancement and low-light enhancement without introducing additional color distortion.

\subsection{Ablation Study}
We conduct extensive ablation studies to verify the effectiveness of different modules in our PUGAN. According to the structure of the model, it is divided into three parts for experimental verification, including a total of 11 designs:
\begin{itemize}
\item[$\bullet$] No.1: `$J^{'}$' denotes the performance evaluation of the output of Par-subnet in Phy-G.
\item[$\bullet$] No.2: `$J^{'*}$' means the color enhanced image estimated by the outputs of the deepened Par-subnet.
\item[$\bullet$] No.3: `$E^{*}$' indicates using $J^{'*}$ and $I$ as inputs of TISE-subnet.
\item[$\bullet$] No.4: `w/o Estimator ($t$)' indicates replacing the transmission estimator with Eq. (\ref{eq2}).
\item[$\bullet$] No.5: `single-stream with $I$' is a single-stream TSIE-subent with only the $I$ as the input of the top encoder branch, which can be as the baseline of Phy-G.
\item[$\bullet$] No.6: `single-stream with $J^{'}$' is a single-stream TSIE-subent with only the $J^{'}$ as the input of the top encoder branch.
\item[$\bullet$] No.7: `w/o DQ' means replacing the DQ module with a simple concatenation operation.
\item[$\bullet$] No.8: `w/o $dif_k$' removes the left part $dif_k$ in the DQ module when generating $w_k$.
\item[$\bullet$] No.9: `w/o $t_k$' removes the right part $t_k$ in the DQ module when generating $w_k$.
\item[$\bullet$] No.10: `w/o ${L_{GAN_1}}$' means the Discriminator1 is removed.
\item[$\bullet$] No.11: `w/o ${L_{GAN_2}}$' indicates the Discriminator2 is removed.
\end{itemize}

\begin{figure}[!t]
	 \centering
	\includegraphics[width=0.50\textwidth]{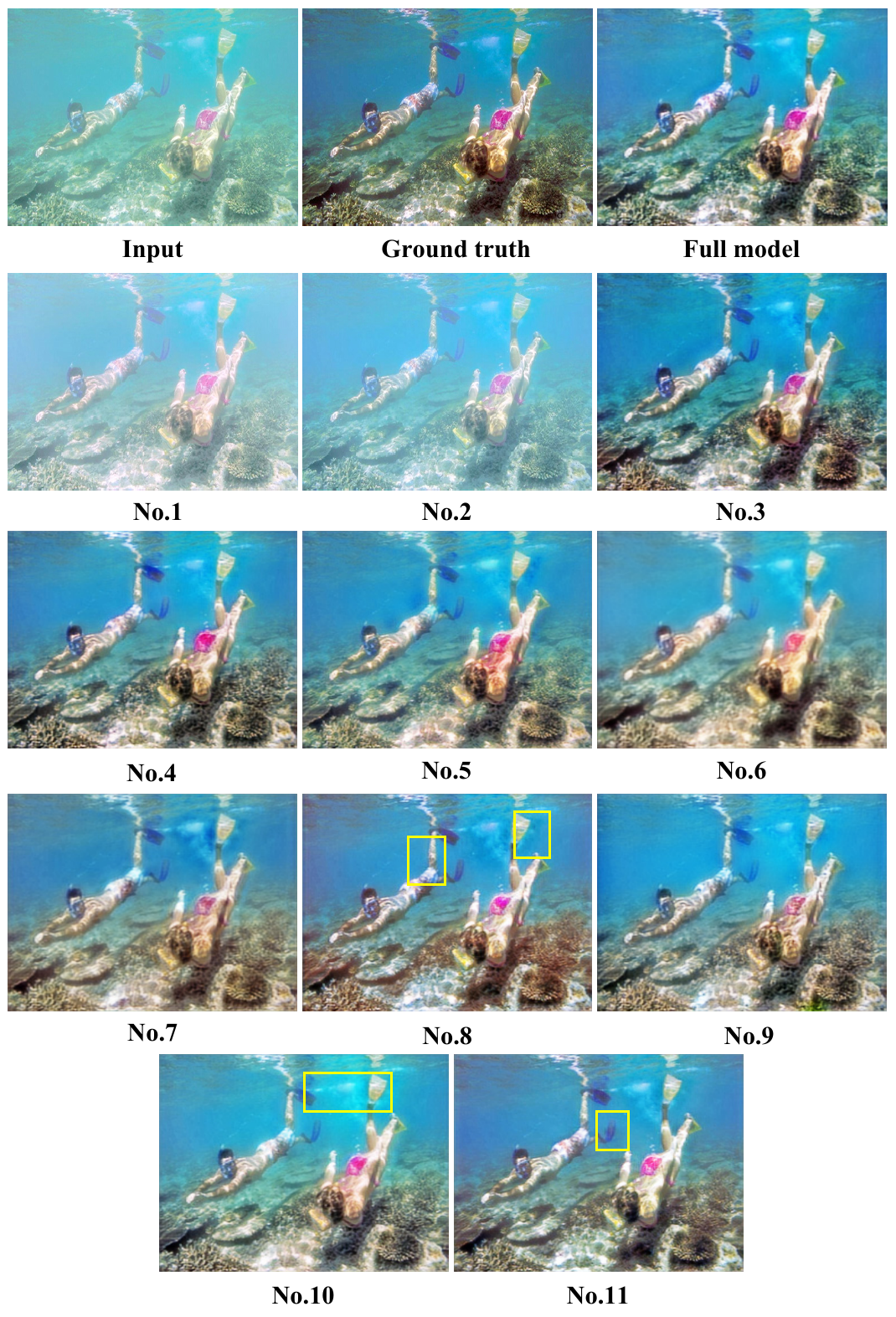}	
	\caption{Visual results of different ablation experiments. }
	\label{fig8}
\end{figure}

As shown in Table \ref{tab3}, the full model outperforms other ablation models in terms of PSNR and MSE scores. We also show the visual results of the ablation experiments in Fig. \ref{fig8}.

\begin{table}[!t]
\centering
\begin{center}
\small
\caption{Quantitative results of the ablation study in terms of average PSNR (DB) and MSE ($\times 10^{3}$) values on the Test-UIEB dataset.} \label{tab3}
\renewcommand\arraystretch{1.2}
\setlength{\tabcolsep}{2.2mm}{ 
\scalebox{0.96}{
\begin{tabular}{c|c|c|c|c}
\hline
\multicolumn{3}{c|}{}&PSNR$\uparrow$ & MSE$\downarrow$ \\ \hline
\multicolumn{3}{c|}{Full model ($E$)} & 21.67      &0.54 \\ \hline
\multirow{4}{*}{{Par-subnet}}  	
& No.1 & $J^{'}$ 	  &18.59      &1.74	 \\ \cline{2-5}
& No.2 & $J^{'*}$      & 19.00      &0.93	 \\ \cline{2-5}
& No.3 & $E^{*}$  	  & 21.48      &0.61\\\cline{2-5}
& No.4 & w/o Estimator ($t$)  	  & 21.08      &0.67\\ \hline
\multirow{5}{*}{{TSIE-subnet}}
& No.5 & single-stream with $I$ 	  &19.87	& 0.77	 \\ \cline{2-5}
& No.6 & single-stream with $J^{'}$	  &20.03     &0.78  \\ \cline{2-5}
& No.7 & w/o DQ	  &19.88      &0.72	 \\ \cline{2-5}
& No.8 & w/o $dif_k$	 &20.71  & 0.68	 \\ \cline{2-5}
& No.9 & w/o $t_k$	  & 20.08      &0.78	 \\ \hline
\multirow{2}{*}{{Dual-D}}
& No.10 & w/o ${L_{GAN_1}}$      & 21.00      &0.60	 \\ \cline{2-5}
& No.11 & w/o ${L_{GAN_2}}$  	  & 20.93      &0.64	\\ \hline
\end{tabular}}}
\end{center}
\end{table}

\textbf{Analysis of the Par-subnet.} 
The Par-subnet in Phy-G aims to generate the color-enhanced image $J^{'}$ through the physical model inversion. According to No.1 reported in Table \ref{tab3} and Fig. \ref{fig8}, the color-enhanced image can achieve effective correction of color distortion, which lives up to our expectations. But it can be found that there is still a big gap between it and our full model, such as the brightness of the enhanced image is too large.

Depth estimation is a pre-operation of transmission map estimation, and its estimation accuracy theoretically does affect some subsequent steps. To evaluate this effect, we design an ablation experiment with different depth estimation network architectures. The original depth estimator uses a relatively shallow network design, including only one RBD module. According to previous experience, network deepening can have a positive impact on estimation accuracy, so we deepen the network by repeating three RBD modules, and the corresponding enhancement results are denoted by asterisk superscripts (\ie, No.2: '$J^{'*}$' and No.3: '$E^{*}$)'.
As reported in No.2 of Table \ref{tab3}, using a more accurate depth map does indeed help improve the performance of the color enhanced image $J^{'}$. For example, after depth estimation network deepening, the MSE is improved from 1.74 (No.1) to 0.93 (No.2). 
This is easy to understand, as there is a relatively direct causal relationship between the depth map and the color enhanced image. However, comparing the $E$ and $E^{*}$, the improvement of depth map performance in the overall enhancement network may not necessarily bring significant improvements to the final results. On the one hand, there is already a relatively weak correlation between the depth map and the final result. On the other hand, the second-stage enhancement process is a nonlinear fitting process based on deep learning, which has good fault tolerance and accuracy. Therefore, non-transformative performance improvements in depth map may not necessarily lead to improvements in the final result. Of course, this also indirectly indicates that our overall model has a certain robustness to depth quality.

In the training process, we give absolute supervision to the attenuation coefficient estimator and depth estimator. Then, the depth map $d_1$ and the attenuation coefficient $\beta$ are used to estimate the transmission map $t$. In fact, we could directly use Eq. (\ref{eq2}) to estimate the transmission map. But considering the stronger modeling capabilities of the learning-based methods, we design a learning-based transmission estimator. However, we do not supervise the estimation of the transmission map with labels. To ensure its quality, we again compute the depth map $d_2$ using Eq. (\ref{eq10}), and then utilize the supervision of the depth map to constrain the learning of the transmission map. Although the ultimate goal of the transmission map estimation is to simulate the calculation method of Eq. (\ref{eq2}) in terms of supervision, learning-based methods are often better suited for nonlinear fitting, which can help to achieve better results. To verify this, we also conduct an ablation experiment to compare the learning-based transmission estimator (No.1) with the Eq. (\ref{eq2})-based transmission map estimation method (No.4). From the results in Table \ref{tab3}, it can be observed that our designed learning-based transmission estimator achieves better performance, which also verifies our previous analysis. 

\textbf{Analysis of the TSIE-subnet.}  
In this part, we first validate the two-stream architecture in the TSIE-subnet and conduct two ablation experiments (\ie, No.5: `single-stream with $I$' and No.6: `single-stream with $J^{'}$'). 
From the quantitative results in Table \ref{tab3}, under the single-stream architecture, the performance of using $J^{'}$ or $I$ is far inferior to the full model of the two-stream architecture. Moreover, from the visualization results in Fig. \ref{fig8}, the enhancement effect of No.5 is relatively blurred, while the result of No.6 has a dispersion phenomenon. In general, the two-stream structure we designed in TSIE-subnet is effective, and introducing $J^{'}$ as guidance information in the TSIE-subnet can further improve the performance, which illustrates the necessity of introducing the physical model guidance.
Subsequently, in order to verify the role of the DQ module in the TSIE-subnet, we design three ablation experiments, including `w/o DQ' (No.7), `w/o $dif_k$' (No.8), and `w/o $t_k$' (No.9). 
In the DQ module design, we jointly consider the feature differences and the transmission characteristics to quantify the seriously degraded regions in the scene, thereby guiding the encoder features fusion and reinforcement.
As shown in No.7 of Table \ref{tab3}, the performance of replacing the DQ module with a simple concatenation is greatly reduced, which shows the effectiveness of our overall architecture design. If we keep only part of the DQ modules (such as No.8 and No.9 in Table \ref{tab3}), the performance is better than removing the DQ modules completely, but not as good as the full model. Moreover, as shown in Fig. \ref{fig8}, only retaining the $t_k$ branch in DQ module will introduce some artifacts, and retaining only $dif_k$ is easy to generate reddish results (such as the coral area in the image).




\textbf{Analysis of the Dual-D.} In our PUGAN, we design the Dual-Discriminators for the style-content adversarial constraint, promoting the authenticity and visual aesthetics of the results. To verify the effectiveness of these two GAN losses, we remove one of them from the full model for ablation experiments. From Table \ref{tab3}, we can see that removing any discriminator in Dual-D negatively affects the quality of results in terms of evaluation metrics and visual quality.  
As shown in Fig. \ref{fig8}, removing the discriminator1 will weaken the ability to correct color distortion, such as the color of the coral area in the lower left corner and the color of women's swimming trunks. Similarly, removing discriminator2 may introduce some new distortions, such as newly generated black artifacts in men's swimming shoes. In contrast, our full model performs well in color correction and content authenticity.



\subsection{Discussions}
Table \ref{tab4} provides a comparison of the performance and model size of different algorithms. It can be seen that our final model achieves the best performance, but its model size reaches 660 MB, and the testing time is 0.14s per image. The overall performance of Ucolor method \cite{li2021underwater} is second only to our method, with a model size of 616 MB. From the model size, our PUGAN is indeed not competitive. During the design process, our model consists of two stages that require parameter estimation and correction enhancement. Moreover, TSIE-subnet is based on UNet, the skip-connections and linear layers in the model consume a lot of computing resources. In the future, we can try to reduce our model parameters in the following ways: choosing a lightweight backbone network and removing some possible redundant parameters in the network, especially the fully connected layers.

\begin{table}[t]
\centering
\renewcommand\arraystretch{1.3}
			\caption{Comparison with the size and performance of the comparison algorithms on the Test-UIEB dataset. }
\scalebox{1.1}{
\begin{tabular}{c|c|c}\hline
\hline
Methods    &Size (MB)       &PSNR$\uparrow$  \\\hline
deep-sesr \cite{islam2020simultaneous}   & 20      & 15.57 \\\hline    
FUnIE-GAN \cite{islam2020fast}      & 218     &18.07\\\hline
WaterNet \cite{li2019underwater}    & 158      & 19.81 \\\hline       
TOPAL \cite{9774330}        & 44       & 19.85    \\\hline           
Ucolor \cite{li2021underwater}     & 616      & 20.61  \\\hline       
PUGAN        & 660         & 21.67            \\\hline
\hline
\end{tabular}
}
\label{tab4}
\end{table}

\begin{figure}[!t]
\centering

\subfigure{
\begin{minipage}[t]{1\linewidth}
\centering
\includegraphics[width=1\textwidth]{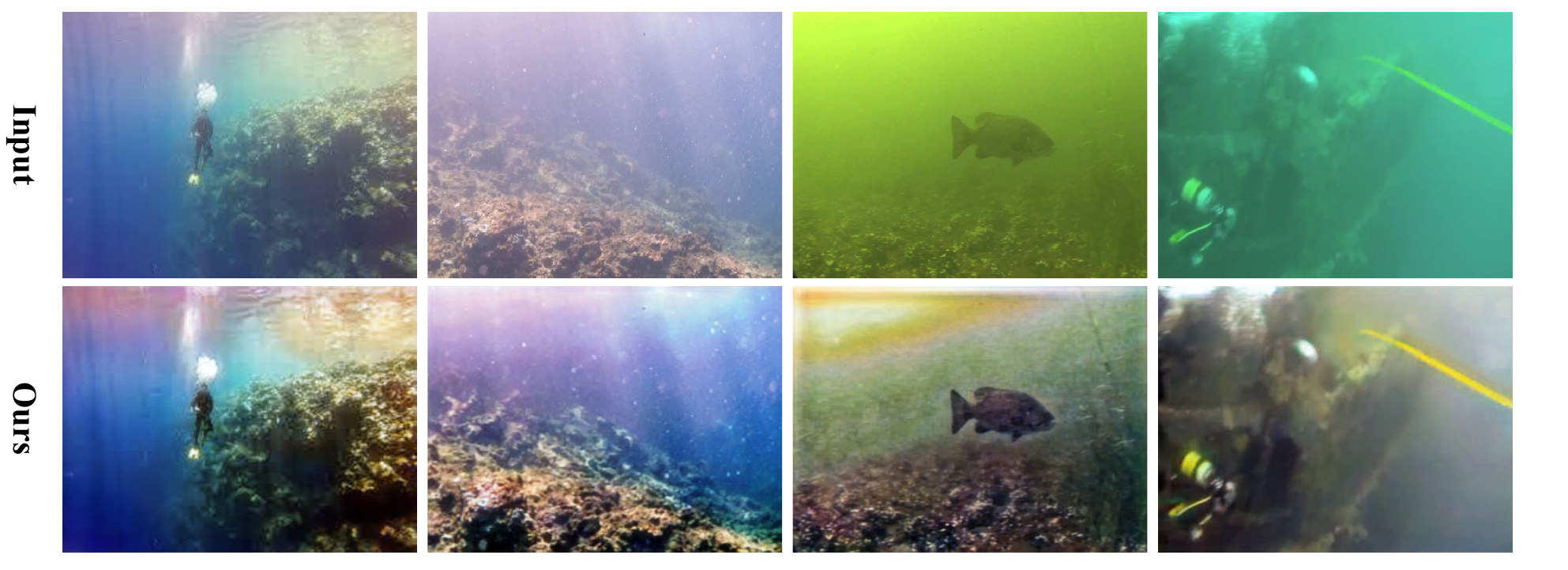}
\end{minipage}
}%
\centering
\caption{Failure Cases. Our method does not work well in some cases where the quality degradation is very severe, such as turbid water, or uneven lighting.}

\label{1}
\label{fig_failure}
\end{figure}

In addition, we discuss some failure cases of our PUGAN, as shown in Fig. \ref{fig_failure}. Our method does not work well in some cases where the quality degradation is very severe, such as turbid water, uneven lighting, \etc. For example, in the second and fourth images, the water quality is relatively turbid, and it is almost difficult to observe its edges, details, and other information. The UIE task at this point is more of a generation task, so the enhancement model may cause unreasonably sudden color changes (\eg, the bottom right area of the second image has an unreasonably green color). Also, uneven lighting can affect our enhancement result. As shown in the third image, the brightness of the upper left area is significantly different from the brightness of the lower area. Since our model does not specifically design the lighting factor, this will make the enhanced results appear to be over-enhanced in areas with relatively high brightness. In the future, it can be considered to introduce edge refinement and brightness correction modules on the basis of ensuring color correction to achieve targeted enhancement of severely blurred and unevenly illuminated images.

\section{CONCLUSION}

  In this paper, we propose a physical model-guided GAN model for underwater image enhancement. In the phy-G, we fully combine the physical model and the CNN-based model, where the Par-subnet generates the color-enhanced underwater image by physical inversion, and the TSIE-subnet equipped with a DQ module aims to generate the final enhanced image through the regional and differential feature learning. In addition, we design a novel Dual-D structure to judge the reconstruction results of the generator, following a style-content synergy mechanism. Our extensive experiments on different benchmarks demonstrate the superiority of this method and the effectiveness of each module.




\bibliographystyle{IEEEtran}
\bibliography{IEEEabrv,Bibliography}

\begin{thebibliography}{10}
\providecommand{\url}[1]{#1}
\csname url@rmstyle\endcsname
\providecommand{\newblock}{\relax}
\providecommand{\bibinfo}[2]{#2}
\providecommand\BIBentrySTDinterwordspacing{\spaceskip=0pt\relax}
\providecommand\BIBentryALTinterwordstretchfactor{4}
\providecommand\BIBentryALTinterwordspacing{\spaceskip=\fontdimen2\font plus
\BIBentryALTinterwordstretchfactor\fontdimen3\font minus
  \fontdimen4\font\relax}
\providecommand\BIBforeignlanguage[2]{{%
\expandafter\ifx\csname l@#1\endcsname\relax
\typeout{** WARNING: IEEEtran.bst: No hyphenation pattern has been}%
\typeout{** loaded for the language `#1'. Using the pattern for}%
\typeout{** the default language instead.}%
\else
\language=\csname l@#1\endcsname
\fi
#2}}

\bibitem{tao2017image}
Z.~Tao, H.~Liu, H.~Fu, and Y.~Fu, ``{Image cosegmentation via saliency-Guided
  constrained clustering with cosine similarity},'' in \emph{Proc. AAAI Conf.
  Artif. Intell. (AAAI)}, vol.~31, no.~1, 2017, pp. 4285--4291.

\bibitem{li2019underwater}
C.~Li, C.~Guo, W.~Ren, R.~Cong, J.~Hou, S.~Kwong, and D.~Tao, ``An underwater
  image enhancement benchmark dataset and beyond,'' \emph{IEEE Trans. Image
  Process.}, vol.~29, pp. 4376--4389, 2019.

\bibitem{hambarde2021uw}
P.~Hambarde, S.~Murala, and A.~Dhall, ``{UW-GAN: Single-image depth estimation
  and image enhancement for underwater images},'' \emph{IEEE Trans. Instrum.
  Meas.}, vol.~70, pp. 1--12, 2021.

\bibitem{peng2018generalization}
Y.-T. Peng, K.~Cao, and P.~C. Cosman, ``Generalization of the dark channel
  prior for single image restoration,'' \emph{IEEE Trans. Image Process.},
  vol.~27, no.~6, pp. 2856--2868, 2018.

\bibitem{islam2020fast}
M.~J. Islam, Y.~Xia, and J.~Sattar, ``{Fast underwater image enhancement for
  improved visual perception},'' \emph{IEEE Robot. Autom. Lett.}, vol.~5,
  no.~2, pp. 3227--3234, 2020.

\bibitem{li2021underwater}
C.~Li, S.~Anwar, J.~Hou, R.~Cong, C.~Guo, and W.~Ren, ``{Underwater image
  enhancement via medium transmission-guided multi-color space embedding},''
  \emph{IEEE Trans. Image Process.}, vol.~30, pp. 4985--5000, 2021.

\bibitem{9825662}
Z.~Huang, J.~Li, Z.~Hua, and L.~Fan, ``Underwater image enhancement via
  adaptive group attention-based multiscale cascade transformer,'' \emph{IEEE
  Trans. Instrum. Meas.}, vol.~71, pp. 1--18, 2022.

\bibitem{guo2020zero}
C.~Guo, C.~Li, J.~Guo, C.~C. Loy, J.~Hou, S.~Kwong, and R.~Cong,
  ``Zero-reference deep curve estimation for low-light image enhancement,'' in
  \emph{Proc. IEEE/CVF Conf. Comput. Vis. Pattern Recognit. (CVPR)}, 2020, pp.
  1777--1786.

\bibitem{li2021learning}
C.~Li, C.~Guo, and C.~C. Loy, ``{Learning to enhance low-light image via
  zero-reference deep curve estimation},'' \emph{IEEE Trans. Pattern Anal.
  Mach. Intell.}, vol.~44, no.~8, pp. 4225--4238, 2022.

\bibitem{crm/tits22/low-light}
Q.~Jiang, Y.~Mao, R.~Cong, W.~Ren, C.~Huang, and F.~Shao, ``Unsupervised
  decomposition and correction network for low-light image enhancement,''
  \emph{IEEE Trans. Intell. Transp. Syst.}, early access, doi:
  10.1109/TITS.2022.3165176.

\bibitem{9447190}
G.~Kim, S.~W. Park, and J.~Kwon, ``Pixel-wise wasserstein autoencoder for
  highly generative dehazing,'' \emph{IEEE Trans. Image Process.}, vol.~30, pp.
  5452--5462, 2021.

\bibitem{9252912}
D.~Zhao, L.~Xu, L.~Ma, J.~Li, and Y.~Yan, ``Pyramid global context network for
  image dehazing,'' \emph{IEEE Trans. Circuits Syst. Video Technol.}, vol.~31,
  no.~8, pp. 3037--3050, 2021.

\bibitem{crm/mtap22/dehazing}
Z.~Wang, F.~Li, R.~Cong, H.~Bai, and Y.~Zhao, ``Adaptive feature fusion network
  based on boosted attention mechanism for single image dehazing,''
  \emph{Multim. Tools Appl.}, vol.~81, no.~8, pp. 11\,325--11\,339, 2022.

\bibitem{crm/tmm20/dehazing}
C.~Li, C.~Guo, J.~Guo, P.~Han, H.~Fu, and R.~Cong, ``{PDR-Net}:
  perception-inspired single image dehazing network with refinement,''
  \emph{{IEEE} Trans. Multim.}, vol.~22, no.~3, pp. 704--716, 2020.

\bibitem{iqbal2010enhancing}
K.~Iqbal, M.~Odetayo, A.~James, R.~A. Salam, and A.~Z.~H. Talib, ``{Enhancing
  the low quality images using unsupervised colour correction method},'' in
  \emph{IEEE International Conference on Systems, Man and Cybernetics}, 2010,
  pp. 1703--1709.

\bibitem{ghani2015underwater}
A.~S.~A. Ghani and N.~A.~M. Isa, ``Underwater image quality enhancement through
  integrated color model with {Rayleigh} distribution,'' \emph{Applied Soft
  Computing}, vol.~27, pp. 219--230, 2015.

\bibitem{ancuti2017color}
C.~O. Ancuti, C.~Ancuti, C.~De~Vleeschouwer, and P.~Bekaert, ``{Color balance
  and fusion for underwater image enhancement},'' \emph{IEEE Trans. Image
  Process.}, vol.~27, no.~1, pp. 379--393, 2018.

\bibitem{gao2019underwater}
S.-B. Gao, M.~Zhang, Q.~Zhao, X.-S. Zhang, and Y.-J. Li, ``{Underwater image
  enhancement using adaptive retinal mechanisms},'' \emph{IEEE Trans. Image
  Process.}, vol.~28, no.~11, pp. 5580--5595, 2019.

\bibitem{ancuti2012enhancing}
C.~Ancuti, C.~O. Ancuti, T.~Haber, and P.~Bekaert, ``{Enhancing underwater
  images and videos by fusion},'' in \emph{Proc. IEEE/CVF Conf. Comput. Vis.
  Pattern Recognit. (CVPR)}, 2012, pp. 81--88.

\bibitem{he2010single}
K.~He, J.~Sun, and X.~Tang, ``{Single image haze removal using dark channel
  prior},'' \emph{IEEE Trans. Pattern Anal. Mach. Intell.}, vol.~33, no.~12,
  pp. 2341--2353, 2010.

\bibitem{galdran2015automatic}
A.~Galdran, D.~Pardo, A.~Pic{\'o}n, and A.~Alvarez-Gila, ``{Automatic
  red-channel underwater image restoration},'' \emph{J. Vis. Commun. Image
  Represent.}, vol.~26, pp. 132--145, 2015.

\bibitem{drews2016underwater}
P.~L. Drews, E.~R. Nascimento, S.~S. Botelho, and M.~F.~M. Campos,
  ``{Underwater depth estimation and image restoration based on single
  images},'' \emph{IEEE Comput. Graph. Appl.}, vol.~36, no.~2, pp. 24--35,
  2016.

\bibitem{crm/JEI16/underwater}
C.~Li, J.~Guo, B.~Wang, R.~Cong, Y.~Zhang, and J.~Wang, ``Single underwater
  image enhancement based on color cast removal and visibility restoration,''
  \emph{J. Electronic Imaging}, vol.~25, no.~3, p. 033012, 2016.

\bibitem{li2016underwater}
C.-Y. Li, J.-C. Guo, R.-M. Cong, Y.-W. Pang, and B.~Wang, ``{Underwater image
  enhancement by dehazing with minimum information loss and histogram
  distribution prior},'' \emph{IEEE Trans. Image Process.}, vol.~25, no.~12,
  pp. 5664--5677, 2016.

\bibitem{zhang2022underwater}
W.~Zhang, Y.~Wang, and C.~Li, ``{Underwater image enhancement by attenuated
  color channel correction and detail preserved contrast enhancement},''
  \emph{IEEE J. Ocean. Eng.}, pp. 1--18, 2022.

\bibitem{crm/tcyb22/glnet}
R.~Cong, N.~Yang, C.~Li, H.~Fu, Y.~Zhao, Q.~Huang, and S.~Kwong,
  ``Global-and-local collaborative learning for co-salient object detection,''
  \emph{IEEE Trans. Cybern.}, vol.~53, no.~3, pp. 1920--1931, 2023.

\bibitem{crm/acmmm21/CDINet}
C.~Zhang, R.~Cong, Q.~Lin, L.~Ma, F.~Li, Y.~Zhao, and S.~Kwong,
  ``Cross-modality discrepant interaction network for {RGB-D} salient object
  detection,'' in \emph{Proc. ACM MM}, 2021, pp. 2094--2102.

\bibitem{crm/nips20/CoADNet}
Q.~Zhang, R.~Cong, J.~Hou, C.~Li, and Y.~Zhao, ``{CoADNet}: Collaborative
  aggregation-and-distribution networks for co-salient object detection,'' in
  \emph{Proc. NeurIPS}, 2020, pp. 6959--6970.

\bibitem{crm/tetci22/PSNet}
R.~Cong, W.~Song, J.~Lei, G.~Yue, Y.~Zhao, and S.~Kwong, ``{PSNet}: Parallel
  symmetric network for video salient object detection,'' \emph{IEEE Trans.
  Emerg. Topics Comput. Intell.}, early access, doi:
  10.1109/TETCI.2022.3220250.

\bibitem{crm/tmm22/TNet}
R.~Cong, K.~Zhang, C.~Zhang, F.~Zheng, Y.~Zhao, Q.~Huang, and S.~Kwong, ``Does
  {Thermal} really always matter for {RGB-T} salient object detection?''
  \emph{IEEE Trans. Multimedia}, early access, doi: 10.1109/TMM.2022.3216476.

\bibitem{crm/tip22/CIRNet}
R.~Cong, Q.~Lin, C.~Zhang, C.~Li, X.~Cao, Q.~Huang, and Y.~Zhao, ``{CIR-Net}:
  Cross-modality interaction and refinement for {RGB-D} salient object
  detection,'' \emph{IEEE Trans. Image Process.}, vol.~31, pp. 6800--6815,
  2022.

\bibitem{crm/tnnls22/360SOD}
R.~Cong, K.~Huang, J.~Lei, Y.~Zhao, Q.~Huang, and S.~Kwong, ``Multi-projection
  fusion and refinement network for salient object detection in 360$^{\circ}$
  omnidirectional image,'' \emph{IEEE Trans. Neural Netw. Learn. Syst.}, early
  access, doi: 10.1109/TNNLS.2022.3233883.

\bibitem{crm/tcsvt22/weaklySOD}
R.~Cong, Q.~Qin, C.~Zhang, Q.~Jiang, S.~Wang, Y.~Zhao, and S.~Kwong, ``A weakly
  supervised learning framework for salient object detection via hybrid
  labels,'' \emph{IEEE Trans. Circuits Syst. Video Technol.}, vol.~33, no.~2,
  pp. 534--548, 2023.

\bibitem{crm/jbhi22/polyp}
G.~Yue, W.~Han, B.~Jiang, T.~Zhou, R.~Cong, and T.~Wang, ``Boundary constraint
  network with cross layer feature integration for polyp segmentation,''
  \emph{IEEE J. Biomed. Health Inform.}, vol.~26, no.~8, pp. 4090--4099, 2022.

\bibitem{crm/tce22/covid}
R.~Cong, Y.~Zhang, N.~Yang, H.~Li, X.~Zhang, R.~Li, Z.~Chen, Y.~Zhao, and
  S.~Kwong, ``Boundary guided semantic larning for real-time {COVID-19} lung
  infection segmentation system,'' \emph{IEEE Trans. Consum. Electron.},
  vol.~68, no.~4, pp. 376--386, 2022.

\bibitem{crm/tip20/MCMT-GAN}
Y.~Huang, F.~Zheng, R.~Cong, W.~Huang, M.~R. Scott, and L.~Shao, ``{MCMT-GAN:}
  multi-task coherent modality transferable {GAN} for 3d brain image
  synthesis,'' \emph{{IEEE} Trans. Image Process.}, vol.~29, pp. 8187--8198,
  2020.

\bibitem{crm/tim22/covid}
R.~Cong, H.~Yang, Q.~Jiang, W.~Gao, H.~Li, C.~Wang, Y.~Zhao, and S.~Kwong,
  ``{BCS-Net}: Boundary, context, and semantic for automatic {COVID-19} lung
  infection segmentation from {CT} images,'' \emph{{IEEE} Trans. Instrum.
  Meas.}, vol.~71, pp. 1--11, 2022.

\bibitem{crm/tcyb22/rsi}
X.~Zhou, K.~Shen, L.~Weng, R.~Cong, B.~Zheng, J.~Zhang, and C.~Yan,
  ``Edge-guided recurrent positioning network for salient object detection in
  optical remote sensing images,'' \emph{IEEE Trans. Cybern.}, vol.~53, no.~1,
  pp. 539--552, 2023.

\bibitem{crm/tip21/DAFNet}
Q.~Zhang \emph{et~al.}, ``Dense attention fluid network for salient object
  detection in optical remote sensing images,'' \emph{IEEE Trans. Image
  Process.}, vol.~30, pp. 1305--1317, 2021.

\bibitem{crm/tgrs22/RRNet}
R.~Cong, Y.~Zhang, L.~Fang, J.~Li, Y.~Zhao, and S.~Kwong, ``{RRNet}: Relational
  reasoning network with parallel multi-scale attention for salient object
  detection in optical remote sensing images,'' \emph{IEEE Trans. Geosci.
  Remote Sens.}, vol.~60, pp. 1558--0644, 2022.

\bibitem{crm/tgrs19/rsi}
C.~Li, R.~Cong, J.~Hou, S.~Zhang, Y.~Qian, and S.~Kwong, ``Nested network with
  two-stream pyramid for salient object detection in optical remote sensing
  images,'' \emph{IEEE Trans. Geosci. Remote Sens.}, vol.~57, no.~11, pp.
  9156--9166, 2019.

\bibitem{crm/ijcai20/SR}
F.~Li, R.~Cong, H.~Bai, and Y.~He, ``Deep interleaved network for image
  super-resolution with asymmetric co-attention,'' in \emph{Proc. IJCAI}, 2020,
  pp. 534--543.

\bibitem{crm/acmmm21/bridgenet}
Q.~Tang, R.~Cong, R.~Sheng, L.~He, D.~Zhang, Y.~Zhao, and S.~Kwong,
  ``Bridgenet: {A} joint learning network of depth map super-resolution and
  monocular depth estimation,'' in \emph{Proc. ACM MM}, 2021, pp. 2148--2157.

\bibitem{zhang2023controlvideo}
Y.~Zhang, Y.~Wei, D.~Jiang, X.~Zhang, W.~Zuo, and Q.~Tian, ``Controlvideo:
  Training-free controllable text-to-video generation,'' \emph{arXiv preprint
  arXiv:2305.13077}, 2023.

\bibitem{li2018emerging}
C.~Li, J.~Guo, and C.~Guo, ``{Emerging from water: underwater image color
  correction based on weakly supervised color transfer},'' \emph{IEEE Signal
  Process. Lett.}, vol.~25, no.~3, pp. 323--327, 2018.

\bibitem{li2017watergan}
J.~Li, K.~A. Skinner, R.~M. Eustice, and M.~Johnson{-}Roberson, ``{WaterGAN:
  Unsupervised generative network to enable real-time color correction of
  monocular underwater images},'' \emph{IEEE Robot. Autom. Lett.}, vol.~3,
  no.~1, pp. 387--394, 2018.

\bibitem{li2020underwater}
C.~Li, S.~Anwar, and F.~Porikli, ``{Underwater scene prior inspired deep
  underwater image and video enhancement},'' \emph{Pattern Recognit.}, vol.~98,
  p. 107038, 2020.

\bibitem{chen2018deep}
Y.-S. Chen, Y.-C. Wang, M.-H. Kao, and Y.-Y. Chuang, ``{Deep photo enhancer:
  unpaired learning for image enhancement from photographs with GANs},'' in
  \emph{Proc. IEEE/CVF Conf. Comput. Vis. Pattern Recognit. (CVPR)}, 2018, pp.
  6306--6314.

\bibitem{ignatov2017dslr}
A.~Ignatov, N.~Kobyshev, R.~Timofte, K.~Vanhoey, and L.~Van~Gool,
  ``{DSLR-quality photos on mobile devices with deep convolutional networks},''
  in \emph{Proc. IEEE Int. Conf. Comput. Vis. (ICCV)}, 2017, pp. 3277--3285.

\bibitem{zhu2017unpaired}
J.-Y. Zhu, T.~Park, P.~Isola, and A.~A. Efros, ``{Unpaired image-to-image
  translation using cycle-consistent adversarial networks},'' in \emph{Proc.
  IEEE Int. Conf. Comput. Vis. (ICCV)}, 2017, pp. 2242--2251.

\bibitem{yang2022generalization}
H.~Yang and W.~E, ``Generalization error of gan from the discriminator’s
  perspective,'' \emph{Research in the Mathematical Sciences}, vol.~9, no.~1,
  p.~8, 2022.

\bibitem{song2021enhancement}
H.~Song, L.~Chang, Z.~Chen, and P.~Ren,
  ``{Enhancement-Registration-Homogenization (ERH): A comprehensive underwater
  visual reconstruction paradigm},'' \emph{IEEE Trans. Pattern Anal. Mach.
  Intell.}, vol.~44, no.~10, pp. 6953--6967, 2022.

\bibitem{ancuti2019color}
C.~O. Ancuti, C.~Ancuti, C.~De~Vleeschouwer, and M.~Sbert, ``{Color channel
  compensation (3C): A fundamental pre-processing step for image
  enhancement},'' \emph{IEEE Trans. Image Process.}, vol.~29, pp. 2653--2665,
  2020.

\bibitem{lin2020attenuation}
Y.~Lin, L.~Shen, Z.~Wang, K.~Wang, and X.~Zhang, ``{Attenuation coefficient
  guided two-stage network for underwater image restoration},'' \emph{IEEE
  Signal Processing Lett.}, vol.~28, pp. 199--203, 2021.

\bibitem{crm/spl21/underwater}
J.~Hu, Q.~Jiang, R.~Cong, W.~Gao, and F.~Shao, ``Two-branch deep neural network
  for underwater image enhancement in {HSV} color space,'' \emph{{IEEE} Signal
  Process. Lett.}, vol.~28, pp. 2152--2156, 2021.

\bibitem{9774330}
Z.~Jiang, Z.~Li, S.~Yang, X.~Fan, and R.~Liu, ``Target oriented perceptual
  adversarial fusion network for underwater image enhancement,'' \emph{IEEE
  Trans. Circuits Syst. Video Technol}, vol.~32, no.~10, pp. 6584--6598, 2022.

\bibitem{wang2021joint}
K.~Wang, L.~Shen, Y.~Lin, M.~Li, and Q.~Zhao, ``{Joint iterative color
  correction and dehazing for underwater image enhancement},'' \emph{IEEE
  Robot. Autom. Lett.}, vol.~6, no.~3, pp. 5121--5128, 2021.

\bibitem{9018379}
S.~C. Raikwar and S.~Tapaswi, ``Lower bound on transmission using non-linear
  bounding function in single image dehazing,'' \emph{IEEE Trans. Image
  Process.}, vol.~29, pp. 4832--4847, 2020.

\bibitem{9088248}
R.~Li, J.~Pan, M.~He, Z.~Li, and J.~Tang, ``Task-oriented network for image
  dehazing,'' \emph{IEEE Trans. Image Process.}, vol.~29, pp. 6523--6534, 2020.

\bibitem{islam2020simultaneous}
M.~J. Islam, P.~Luo, and J.~Sattar, ``{Simultaneous enhancement and
  super-resolution of underwater imagery for improved visual perception},''
  \emph{arXiv preprint arXiv:2002.01155}, 2020.

\bibitem{li2020deep}
Y.~Li, Y.~Liu, Q.~Yan, and K.~Zhang, ``{Deep dehazing network with latent
  ensembling architecture and adversarial learning},'' \emph{IEEE Trans. Image
  Process.}, vol.~30, pp. 1354--1368, 2021.

\bibitem{isola2017image}
P.~Isola, J.-Y. Zhu, T.~Zhou, and A.~A. Efros, ``{Image-to-image translation
  with conditional adversarial networks},'' in \emph{Proc. IEEE/CVF Conf.
  Comput. Vis. Pattern Recognit. (CVPR)}, 2017, pp. 5967--5976.

\bibitem{9646904}
R.~Chen, Z.~Cai, and W.~Cao, ``{MFFN}: An underwater sensing scene image
  enhancement method based on multiscale feature fusion network,'' \emph{IEEE
  Trans. Geosci. Remote Sens.}, vol.~60, pp. 1--12, 2022.

\bibitem{panetta2015human}
K.~Panetta, C.~Gao, and S.~Agaian, ``{Human-visual-system-inspired underwater
  image quality measures},'' \emph{IEEE J. Ocean. Eng.}, vol.~41, no.~3, pp.
  541--551, 2015.

\bibitem{yang2021reference}
N.~Yang, Q.~Zhong, K.~Li, R.~Cong, Y.~Zhao, and S.~Kwong, ``{A reference-free
  underwater image quality assessment metric in frequency domain},''
  \emph{Signal Processing: Image Communication}, vol.~94, 2021.

\bibitem{7300447}
M.~Yang and A.~Sowmya, ``An underwater color image quality evaluation metric,''
  \emph{IEEE Trans. Image Process.}, vol.~24, no.~12, pp. 6062--6071, 2015.

\bibitem{wang2018imaging}
Y.~Wang, N.~Li, Z.~Li, Z.~Gu, H.~Zheng, B.~Zheng, and M.~Sun, ``An
  imaging-inspired no-reference underwater color image quality assessment
  metric,'' \emph{Computers \& Electrical Engineering}, vol.~70, pp. 904--913,
  2018.

\bibitem{9854113}
P.~Zhuang, J.~Wu, F.~Porikli, and C.~Li, ``Underwater image enhancement with
  hyper-laplacian reflectance priors,'' \emph{IEEE Trans. Image Process.},
  vol.~31, pp. 5442--5455, 2022.

\bibitem{zhang2022underwater11}
W.~Zhang, P.~Zhuang, H.-H. Sun, G.~Li, S.~Kwong, and C.~Li, ``Underwater image
  enhancement via minimal color loss and locally adaptive contrast
  enhancement,'' \emph{IEEE Trans. Image Process.}, vol.~31, pp. 3997--4010,
  2022.

\bibitem{9548907}
J.~Xie, G.~Hou, G.~Wang, and Z.~Pan, ``A variational framework for underwater
  image dehazing and deblurring,'' \emph{IEEE Trans. Circuits Syst. Video
  Technol}, vol.~32, no.~6, pp. 3514--3526, 2022.

\bibitem{9895452}
Y.~Kang, Q.~Jiang, C.~Li, W.~Ren, H.~Liu, and P.~Wang, ``A perception-aware
  decomposition and fusion framework for underwater image enhancement,''
  \emph{IEEE Trans. Circuits Syst. Video Technol}, vol.~22, no.~3, pp.
  988--1002, 2023.

\end{thebibliography}

\vfill


\end{document}